\documentclass[12pt,preprint]{aastex}
\usepackage{amsmath}
\usepackage{color}
\newcommand{\pdel}[2]%
{\frac{\partial{#2}}{\partial{#1}}}
\newcommand{\pddel}[2]%
{\frac{\partial^2{#2}}{\partial{#1}^2}}
\newcommand{\udel}[1]{\partial_{#1}}
\newcommand{\tdel}[1]{\partial^{#1}}

\newcommand{\Ms}{M_{\odot}}
\newcommand{\km}{\mathrm{km}}
\newcommand{\cm}{\mathrm{cm}}

\newcommand{\cmps}{\mathrm{cm/s}}

\newcommand{\gauss}{\mathrm{G}}
\newcommand{\erg}{\mathrm{erg}}
\newcommand{\psec}{\mathrm{/s}}
\newcommand{\gpcmc}{\mathrm{g/cm^3}}
\newcommand{\pcmc}{\mathrm{/cm^3}}
\newcommand{\pstr}{\mathrm{/str}}

\newcommand{\gpccm}{{ \rm g \,\, cm^{-3}}}

\slugcomment{Accepted to the ApJ}
\shorttitle{Long-Term Evolution of Collapsar}
\shortauthors{Harikae et al.}

\begin{document}

\title{
 Long-Term Evolution of Slowly Rotating Collapsar 
in Special Relativistic Magnetohydrodynamics}

\author{Seiji Harikae\altaffilmark{1,2},
Tomoya Takiwaki\altaffilmark{3},
 and Kei Kotake\altaffilmark{2,3}}

\affil{\altaffilmark{1}Department of Astronomy, The Graduate School of Science, 
University of Tokyo, 7-3-1 Hongo, Bunkyo-ku,Tokyo, 113-0033, Japan}
\affil{\altaffilmark{2}Division of Theoretical Astronomy, 
  National Astronomical Observatory of Japan, 2-21-1, 
  Osawa, Mitaka, Tokyo, 181-8588, Japan}
\affil{\altaffilmark{3}Center for Computational Astrophysics,
 National Astronomical Observatory of Japan, 2-21-1, 
  Osawa, Mitaka, Tokyo, 181-8588, Japan}


\email{seiji.harikae@nao.ac.jp}

\begin{abstract}
We present our numerical results of 
 two-dimensional magnetohydrodynamic (MHD) simulations 
of the collapse of rotating massive stars 
in light of the collapsar model of gamma-ray bursts (GRBs).
Pushed by recent evolution calculations of
  GRB progenitors, we focus on lower angular momentum of the central 
core than the ones taken mostly in previous studies.
  By performing special relativistic simulations including 
both realistic equation of state and neutrino cooling, we follow a long-term evolution of the slowly rotating collapsars up to $\sim$ 10 s,
accompanied by the formation of jets and accretion disks. 
Our results show that for the GRB progenitors to function as collapsars, 
there is a critical initial angular momentum, below which 
matter is quickly swallowed to the central objects, no accretion disks and no
 MHD outflows are formed.
When larger than the criteria, 
 we find the launch of the MHD jets in the following two ways.
 For models with stronger initial magnetic fields, the magnetic 
pressure amplified inside the accretion disk 
 can drive the MHD outflows, which makes the strong magnetic explosions 
 like a 'magnetic tower'. For models with weaker 
 initial magnetic fields, the magnetic tower stalls first and the subsequent MHD 
outflows are produced by the magnetic twisting of the turbulent inflows 
of the accreting material from the equatorial to the polar regions.
Regardless of the difference in the formation,  the jets can 
 attain only mildly relativistic speeds with 
the explosion energy less than $10^{49} \erg$.
 To obtain stronger neutrino energy depositions 
in the polar funnel regions 
heated from the accretion disk, we find that 
 smaller initial angular momentum is favorable. This is 
because the gravitational compression makes 
 the temperature of the disk higher.
Due to high neutrino opacity inside the disk, 
we find that the luminosities of 
 $\nu_e$ and $\bar{\nu}_e$ become almost comparable, which is advantageous for 
 making the energy deposition rate larger. 
We discuss how the energy deposition
 can be as efficient as the magnetically-driven 
processes for energetizing jets.
 Among the 
computed models, we suggest that the model with the initial angular momentum 
of $j \sim 1.5j_{\mathrm{lso}}$ ($j_{\mathrm{lso}}$: the angular momentum of 
the last stable orbit) and with initial magnetic field strength 
 of $\sim 10^{10}$ G, provides a most plausible condition for making 
 fireballs for GRBs,
 because such model is appropriate not only for producing the MHD outflows 
 quickly by the magnetic towers, but also for obtaining the 
 stronger neutrino heating in the evacuated polar funnel.
\end{abstract}


\section{Introduction}
Gamma-ray bursts (GRBs) are one of the most energetic phenomena
 in the universe. Accumulating observations discovered such as by {\it Swift} {\footnote {http://www.swift.psu.edu/}}
 and {\it HETE-2} {\footnote {http://space.mit.edu/HETE/}} show that GRBs are basically categorized into two, namely short-hard and long-soft bursts
(e.g., \citet{nakar07} for review). 
More surprisingly, GRBs with some 
mixed features of the two types have been reported
(e.g., \citet{gehre06}, \citet{galy06}). The mystery of their central engines 
 seems to be thickening, which have long puzzled astrophysicists
 since the accidental discovery in the late sixties 
(e.g., \citet{mesz06}, for review).
Speaking about the long-duration GRBs, a number of their host galaxies have been
 identified as metal poor galaxies 
\citep[][and reference therein]{sava06,stan06}.
The preponderance of short-lived massive star formation in such
young galaxies, as well as the identification of SN Ib/c light curves
 in some bursts, has provided strong support to identify 
a massive stellar collapse as an origin of the long GRBs 
\citep{pacz98,gala98,hjor03,
stan03}. The duration of the long bursts 
may correspond to the accretion of debris falling into the 
central black
hole (BH)\citep{piro98}, which suggests the observational consequence
of the BH formation likewise the supernova of neutron star formation.
Pushed by those observations, the so-called collapsar 
 has received quite some interest for the central engines of the 
 long GRBs \citep{woos93,pacz98,macf99}.

 In the collapsar model, the central cores 
with significant angular momentum collapse into a BH. 
Neutrinos emitted from the accretion disk around the BH heat the matter of the 
funnel region of the disk, to
launch the GRB outflows. The relativistic outflows are expected to ultimately 
form a fireball, which is 
good for explaining the observed afterglow (e.g.,  \citet{pira99}).
In addition, it is suggested that the strong magnetic fields in the
cores of order of $10^{15} \gauss$ play also an active role both for driving 
the magneto-driven jets and for extracting a significant amount of
energy from the central engine (e.g.,
\cite{whee00,thom04,uzde07a} and see references therein). 

To obtain clear understanding of such scenarios, it is ultimately 
 necessary to perform stellar core-collapse simulations, which trace
all the phases in a consistent manner starting from the stellar core-collapse, 
core-bounce, shock-stall, stellar explosion (phase 1) or BH formation
and the formation of accretion disk (phase 2), energy
deposition to the funnel region by neutrinos and/or magnetic fields
(phase 3), to the launching of the fireballs (phase 4). 
Here for convenience we call each stage as phase 1, 2, etc.
The requirement for the numerical modeling to this end is highly
computationally expensive, which necessitates the multidimensional 
magnetohydrodynamic(MHD) simulations not only with general relativity for handling the BH
formation, but also with the multi-angle neutrino transfer for
treating highly anisotropic neutrino radiation from the disks.
Thus various approximate approaches have been undertaken.
All the studies, which we will mention below, are complimentary in the sense that 
the different epochs are focused on, with the different initial
conditions for the numerical modeling being taken.

As for the phase 1, the roles of rapid rotation and magnetic fields have been 
 elaborately investigated to study 
the formation of magnetars and its implications to
the collapsars (e.g., \citet{taki04,kota04a,sawa05,ober06a,suwa07a,burr07,taki09} and 
collective references in \citet{kotake_rev}).
After the failed or weak explosion in the postbounce phase,
the accretion to the central objects may lead to the formation of a BH (phase 2), 
 which several general relativistic studies have focused on \citep{shib06,seki07}. 
Treating the BH as an absorbing boundary or using the fixed metric
approaches, the numerical studies of 
the phase 3 are concerned with the initiation of the outflows from the funnel
region of the disk to the acceleration of the jets as a result of the
neutrino heating and/or MHD processes till the jets become mildly relativistic. 
Numerical studies of the phase 4 are 
mainly concerned with the dynamics later on, namely, the jet
propagation to the breakout from the star, when the acceleration of
the jets to the high Lorentz factor is expected (see, e.g., 
\citet{aloy00,zhan03,mizu07} and references therein).


Here we focus on the phase 3, which has also been extensively investigated thus 
far (e.g., \citet{macf99,prog03c,devi05,hawl06,mizu06,fuji06,mcki07b,komi07a,naga07,bark08}). 
A general outcome of the MHD studies among them 
(e.g., \citet{prog03b,prog03c,mizu06,fuji06,naga07,naga09})
is that if the central progenitor cores 
 have significant angular momentum ($\approx 10^{17}$ $\rm {cm}^2$/s) 
with strong magnetic fields ($\gtrsim 10^{11}$ G), 
magneto-driven jets can be launched strong 
enough to expel the matter along the rotational axis within several seconds after the 
onset of collapse. 
The combination of such 
rapid rotation and strong magnetic fields 
 is recently considered to be possible for rapidly rotating metal-poor stars, 
 which experience the so-called chemically homogeneous evolution in the main sequence 
\citep{woos06,yoon05}.
  It should be noted that the angular momentum of those GRB progenitors 
($\approx 10^{16}$ $\rm {cm}^2$/s), albeit not a final answer due to 
 much uncertainty in the stellar mass loss, angular momentum transport, and 
magnetorotational instability (MRI)
(e.g., \citet{vink05}, \citet{detmers08}), 
is relatively smaller than those assumed in most of the 
previous collapsar simulations.
 Employing such a GRB progenitor, it was reported by
 \citet{dess08} based on the 2D radiation MHD core-collapse simulations that 
 too much angular momentum is not favorable for collapsars because the 
MHD explosions in the immediate postbounce phase are so strong that the BH 
formations are circumvented.

These situations motivate us to focus on slower rotation of the central core
in the context of collapsar models.
As for the initial magnetic fields, we choose to explore relatively smaller 
fields ($\lesssim 10^{10}$ G), which has been less investigated so far.
 Paying particular attention to the smaller angular momentum, it takes much 
longer time to amplify the magnetic fields large enough to launch the MHD jets than 
 previously estimated. By performing special relativistic MHD simulations, 
 which enable us to follow a long-term evolution over $\sim$ 10 s, 
 we aim to clarify how the properties and the mechanism of the MHD jets could 
change with the initial angular momentum and the initial magnetic fields. 
It is noted here that the long-term simulation could be 
 important for understanding the X-ray flares recently discovered 
in a number of long GRBs (e.g., \citet{prog06}).
  As for the microphysics, we include both realistic equation of state (EOS)
  and neutrino cooling, which have been often 
 neglected or oversimplified (see, however, \citet{fuji06,naga07}).
 By doing so, we estimate the neutrino luminosities emitted from the accretion disks, 
 and clarify what conditions are pivotal to make the energy depositions via neutrino 
pair annhilation as efficient as the magnetically-driven 
processes for energetizing jets.
 The range of specific angular momentum required for the progenitors of collapsars 
was predicted to be 
 $3 < j_{16}(\equiv j/10^{16}{\rm cm}^2 ~{\rm s}^{-1}) < 20$ by 
 a pioneering collapsar simulation but without MHD
 \citep{macf99}. This is because if angular momentum is too small, 
mass element cannot stay at the last stable orbit, 
while if too large, mass element cannot fall into compact objects, and thus 
 cannot form disk, suppressing the sufficient energy release for GRBs. 
Based on our results, we hope to understand how this 
 criterion would change if MHD effects are taken into account.

We summarize the numerical methods in \S\ref{sec:num}.
\S 3 is devoted to the initial models.
  The main results are described in \S\ref{sec:results}.
 We summarize our results 
and discuss their implications in  \S\ref{sec:discussion}. 
Details of the treatments of neutrino cooling/heating 
in the framework of special relativity are given in the appendices.

\section{Numerical methods}

\label{sec:num}
The results presented in this paper are calculated by the MHD code 
in special relativity (SR) developed by \citet{taki09}.
In the following, we briefly mention the importance of SR for collapsar simulations
 and summarize shortly the numerical schemes.  

SR effects are indispensable not only to follow the propagation of GRB jets 
in high Lorentz factors, but also to follow the dynamics of infalling material, 
because their free-fall velocities and rotational velocities become close to the 
speed of light near the central compact objects. Moreover,
 the velocity of the Alfv\'{e}n waves during the jet propagation 
can be estimated as,
\begin{equation}
v_{A} = \frac{B}{\sqrt{4 \pi \rho}} 
\sim 10^{10}\mathrm{cm/s}\frac{B/{10^{13}
\mathrm{G}}}{\sqrt{\rho/\left(10^5\gpcmc
\right)}},
\end{equation}
where $\rho$ and $B$ are the typical density and the magnetic field
 near along the rotational axis. 
It can be readily inferred that the Alfv\'{e}n velocity 
can exceed the speed of light unphysically in the Newtonian simulation, 
especially for the regions where the density becomes low and the magnetic fields 
become strong. Such a situation is ubiquitous for collapsar simulations.
Even if the propagation speeds of the jets are only mildly relativistic, 
we have learned that 
 (at least) SR treatments are quite important 
for keeping the stable numerical calculations in good accuracy over the longer-term 
evolution.

The SRMHD part of the code is based on the formalism of \citet{devi03}(see
 \citet{taki09} for details).
To formalize the basic equations, we need two frames; the laboratory frame (we 
 call shortly as lab frame),
 which is the center of mass system of the star,
 and the rest frame of the relativistic fluid. 
These two frames are related to each other with the usual Lorentz transformation. 
Before going to the basic equations, 
we write down the definition of the
primary code variables.
The state of the relativistic fluid element in the rest frame
is described by its density, $\rho$; specific energy, $e$; velocity,
$v^i$; and pressure, $p$.
Magnetic fields in the lab flame is described by
the 4-vector $\sqrt{4\pi}b^{\mu}={^*F}^{\mu\nu}U_{\nu}$, where $^*F^{\mu\nu}$ is the
dual tensor of the electro-magnetic field strength tensor and $U_{\nu}$ is the
4-velocity.

The basic equations of the SRMHD code can be described as follows:
\begin{eqnarray}
\pdel{t}{D}
+\frac{1}{\sqrt{\gamma}}\udel{i}{\sqrt{\gamma}Dv^i} &=&0 \label{eq:mass_consv}\\
\pdel{t}{E}
+\frac{1}{\sqrt{\gamma}}\udel{i}{\sqrt{\gamma}Ev^i}
&=&-p\pdel{t}{W}
-\frac{p}{\sqrt{\gamma}}\udel{i}{\sqrt{\gamma}W v^i}
-{\cal L}_\nu \label{eq:ene_consv}\\
\pdel{t}{(S_i-b^tb_i)}
+\frac{1}{\sqrt{\gamma}}
\udel{j}{\sqrt{\gamma}\left(S_i v^j-b_ib^j\right)}
&=&
-\frac{1}{2}
\left(
\rho h \left(Wv_k\right)^2
- \left(b_k\right)^2
\right)\udel{i}{\gamma^{kk}}\nonumber\\
& &
-\left(\rho h W^2- {b^t}^2\right) \udel{i}{\Phi_\mathrm{tot}}\nonumber\\
& &
-\udel{i}{\left(p+\frac{\|b\|^2}{2}\right)}\label{eq:mom_consv}\\
\pdel{t}{(Wb^i - Wb^tv^i)}
+\udel{j}{\left(Wv^jb^i-Wv^ib^j\right)}
&=&0\label{eq:induction}\\
\tdel{k}{{\udel{k}{\Phi_\mathrm{poi}}}}&=&4\pi \biggl[
\rho h (W^2+(v_k)^2) 
+ 2\left( p+\frac{\left|b\right|^2}{2}\right) \nonumber \\
&&
-\left((b^{t})^2+(b_{k})^2 \right) \biggl]
\label{eq:poisson}
\end{eqnarray}
where $W=\frac{1}{\sqrt{1-v^kv_k}}$, $D=\rho W$, $E=e W$ and 
$S_i=\rho hW^2v_i$ are 
 the Lorentz boost factor, auxiliary variables correspond to density,
energy, and momentum, respectively. 
 All of them are defined in the lab frame. 
Equations.(\ref{eq:mass_consv},\ref{eq:ene_consv},\ref{eq:mom_consv})
represent the mass, energy, and momentum conservations.
In Eq.(\ref{eq:mom_consv}) it is noted that the relativistic enthalpy,
$h=(1+e/\rho+p/\rho+\left|b\right|^2/\rho)$
 includes magnetic energy.
Eq.(\ref{eq:induction}) is the induction equation for the magnetic
fields. In solving the equation, the method of characteristics
is implemented to propagate accurately all modes of MHD waves 
(see \citet{taki09} for details).
$B^i$ is the magnetic field in the rest frame, 
which is related to the one in the lab frame 
as $B^i=Wb^i -Wb_t b^i$.
Here $b_t$ is a time component of the 4-vector of $b_{\mu}$.
Eq.(\ref{eq:poisson}) is the Poisson equation for the 
(self-)gravitational potential of $\Phi_\mathrm{poi}$. 
We employ the realistic equation of state 
based on the relativistic mean field theory \citep{shen98}.
For lower density regime ($\rho \lesssim 10^{5.5} \gpccm$), 
where no data is available in the EOS table with the Shen EOS, 
we use another EOS, which includes contributions
from an ideal gas of nuclei, radiation, and electrons and positrons
with arbitrary degrees of degeneracy~\citep{bdn96}.
We carefully connect two EOS at $\rho = 10^{5.5} \gpccm$
for physical quantities to vary continuous in density at a given temperature.

${\cal L}_\nu$ in Eq.(3) is the neutrino luminosity 
evaluated with a multi-flavor leakage scheme 
(\citet{epst81,ross03,kota03a}), in which 
{$\nu_{e}$, }$\bar{\nu}_{e}$, and
the heavy-lepton neutrinos, $\nu_{\mu}$, $\bar{\nu}_{\mu}$,
$\nu_{\tau}$, $\bar{\nu}_{\tau}$ (collectively referred to
as $\nu_{X}$) are taken into account. 
Electron capture on proton and free nuclei, 
positron capture on neutron, 
photo-pair, plasma process are included 
(\citet{full85}, \citet{taka78}, Itoh et al. 1989, 1990).
Furthermore, we update the leakage scheme to include special relativistic corrections 
 for the first time to our knowledge (see \ref{leak} for details). 

Spherical coordinates, $(r, \theta, \phi)$ are used 
in our simulations and the computational domain is extended over
$50 \km \le r \le 30 000 \km$ and $0 \le \theta \le \pi/2$
and covered with 300($r$) $\times$ 40($\theta$) meshes with the assumption 
 of axial and equatorial symmetry.
As in \citet{macf99,fuji06}, we adopt an absorptive inner boundary condition at 50 km. 
Together with SR treatment of MHD, this inner boundary, albeit taken to 
be large, allows us to explore the long-term evolution of collapsars. 
Note that we do not necessarily assume the formation of BH inside, 
but only assume that the central region would not 
affect the regions outside. 
 One interpretation of the position of the 
inner-boundary could be a surface of the standing accretion shock waves 
produced after bounce.  
In \S \ref{subsec:kerr}, by counting the accreted mass in the central objects 
 when the simulations terminate, 
we will discuss what they could be, namely the neutron stars or the BHs.

The total gravitational potential in Eq. (4) is
\begin{equation}
	\Phi_\mathrm{tot} = \Phi_\mathrm{PW} + \Phi_\mathrm{poi},
\end{equation}
where $\Phi_\mathrm{PW}$ mimics the contribution from 
strong gravity around the central objects \citep{pacz80}.  
Under the special relativistic modification, 
  this potential has been suggested as useful to reproduce 
 the dynamics outside the last stable orbit in the Schwarzschild metric
(\citet{abram96,fukue04}). Thus such treatment, albeit very approximate,
 may not be too bad for our computations (see discussions in \S \ref{subsec:kerr}).
To achieve further accuracy, 
we need to perform simulations in curved time-space 
with general relativistic MHD, which is beyond the scope of this paper.

\section{Initial conditions}\label{sec:inimodel}

\begin{table}[tb]
\begin{center}
\begin{tabular}[c]{cccccc}
\hline
Model& $B_0$& $\alpha$& $T/|W|$& $E_\mathrm{mag}/|W|$\\
\hline
B10J1.0& $10^{10}$G& 1.0& $5.515\times10^{-3}$& $2.108\times10^{-8}$\\
B10J1.5& $10^{10}$G& 1.5& $1.241\times10^{-2}$& $2.108\times10^{-8}$\\
B10J2.0& $10^{10}$G& 2.0& $2.206\times10^{-2}$& $2.108\times10^{-8}$\\
B10J2.5& $10^{10}$G& 2.5& $3.447\times10^{-2}$& $2.108\times10^{-8}$\\
B10J3.0& $10^{10}$G& 3.0& $4.964\times10^{-2}$& $2.108\times10^{-8}$\\
B9J1.0& $10^{9}$G& 1.0& $5.515\times10^{-3}$& $2.108\times10^{-10}$\\
B9J1.5& $10^{9}$G& 1.5& $1.241\times10^{-2}$& $2.108\times10^{-10}$\\
B9J2.0& $10^{9}$G& 2.0& $2.206\times10^{-2}$& $2.108\times10^{-10}$\\
B9J2.5& $10^{9}$G& 2.5& $3.447\times10^{-2}$& $2.108\times10^{-10}$\\
B9J3.0& $10^{9}$G& 3.0& $4.964\times10^{-2}$& $2.108\times10^{-10}$\\
B8J1.0& $10^{8}$G& 1.0& $5.515\times10^{-3}$& $2.108\times10^{-12}$\\
B8J1.5& $10^{8}$G& 1.5& $1.241\times10^{-2}$& $2.108\times10^{-12}$\\
B8J2.0& $10^{8}$G& 2.0& $2.206\times10^{-2}$& $2.108\times10^{-12}$\\
B8J2.5& $10^{8}$G& 2.5& $3.447\times10^{-2}$& $2.108\times10^{-12}$\\
B8J3.0& $10^{8}$G& 3.0& $4.964\times10^{-2}$& $2.108\times10^{-12}$\\
\hline
\end{tabular}
\caption{Models and Parameters. 
Model names are labeled by the initial strength of magnetic fields and rotation.
$B_0$ is a constant in Eq.(\ref{vec_phi}), 
$\alpha$ is the ratio of the specific angular momentum normalized by the 
 one at the last stable orbit in Eq.(\ref{eq:iniang}).
$T/|W|$ and $E_\mathrm{mag}/|W|$ represents the ratio of the rotational energy 
 and the magnetic energy to the absolute value of
  the gravitational energy, respectively. }\label{Table:1}
\end{center}
\end{table}


As for the initial profiles of the collapsing star, 
we employ the spherical data set of the density, temperature, internal energy, 
and electron fraction in model 35OC with the mentioned chemically homogeneous 
evolution \citep{woos06}.  At the the zero age main-sequence,
 the progenitor mass,  rotational velocity, and metallicity are 
$35 \Ms$, $v_\phi = 380 \km/{\rm s}$, and $0.1 Z_{\odot}$, 
respectively. 
At the presupernova phase, the stellar mass is striped to be $28.07 \Ms$ due to 
the mass-loss and the mass of the central iron core is $2.02 \Ms$.
Our numerical grid contains inner $8.56 \Ms$ of the star. 

Since little is known about the spatial configurations of the rotation and
the magnetic fields from the stellar evolution calculations assuming spherical symmetry
 of the stars, we add the following 
rotation and magnetic field profiles in a parametric manner to the core 
mentioned above.

As for the initial angular momentum of the core,
 we parametrize the strength by
 the angular momentum of the last stable orbit (: $j_\mathrm{lso}$)  following 
\citet{lee06,lopez09,prog03a,prog03c,prog03b,prog05} as 
\begin{equation}
j =\alpha j_\mathrm{lso}(M(R)), \label{eq:iniang}
\end{equation}
where $j$ is the specific angular momentum, 
$M(R)$ is the spherical mass coordinate, encompassing the mass inside radius $R$, 
and $\alpha$ is a model parameter.
In this study, we set $1\le\alpha\le3$. 
Note that, although angular momentum is larger 
than $j_\mathrm{lso}$ with this range, 
it does not assure the formation of the stable disk 
because of the existence of the slowly rotating matter 
in the polar regions and relatively large inner boundary of our models. 
Thus dynamical simulations are necessary to specify the criteria of the disk formation.

As for the initial configuration of the magnetic fields, 
 we assume that the poloidal field is nearly uniform and parallel to the rotational 
axis inside the core and dipolar outside.
For the purpose, we consider the following effective vector potential,
\begin{equation}
A_r=A_\theta=0,
\end{equation}
\begin{equation}
 A_\phi=\frac{B_0}{2}\frac{r_0^3}{r^3+r_0^3}r\sin\theta,\label{vec_phi}
\end{equation}
where $A_{r,\theta,\phi}$ is the vector potential in the $r,\theta,\phi$ 
direction, respectively,  $r$ is the radius, $r_0$ is the radius of the core,
 and 
$B_0$ is the model constant. 
We set $r_0 = 3000 \ \km$ between the iron core and the silicon layers 
and change parametrically 
$B_0$ as $B_0 = 10^8, 10^9$ and $10^{10}\ \gauss$ for each model. 

We compute 15 models changing
the initial angular momentum and the strength of magnetic fields
by varying the value of $\alpha$ and $B_{0}$.
Each models are named as BXJY, where X indicates the initial 
poloidal magnetic field ($10^{X}$ G), 
and Y represents the ratio of the specific angular momentum to $j_\mathrm{lso}$. 
For example, B9J1.5 represents the 
model with 
$B_0 = 10^9$ and $j =1.5j_{\mathrm{lso}}$. 
The model parameters are shown in Table \ref{Table:1}.
It is noted that $T/|W|$ and $E_{\rm mag}/|W|$ for the original progenitor of 
 the model 35OC is $2 \times 10^{-3}$ and $1 \times 10^{-6}$, respectively.
Thus the model series with J1.0 have almost the same angular momentum with the 
 progenitor. Considering the mentioned uncertainties of the progenitor models, we 
choose to explore relatively smaller field strength, 
which has been less investigated so far.

\section{Results}
\label{sec:results}
Computing 15 models in a longer time stretch than ever among previous collapsar models,
we observe a wide variety of the dynamics changing drastically with time.
To capture the general properties of all the models, we first pay attention to the 
time evolutions of the central mass, the mass of the accretion disk, 
and the neutrino luminosity in the following.

\subsection{General features}
\subsubsection{Central mass and disk mass}
\begin{figure}[tb]
\epsscale{1.0}\plotone{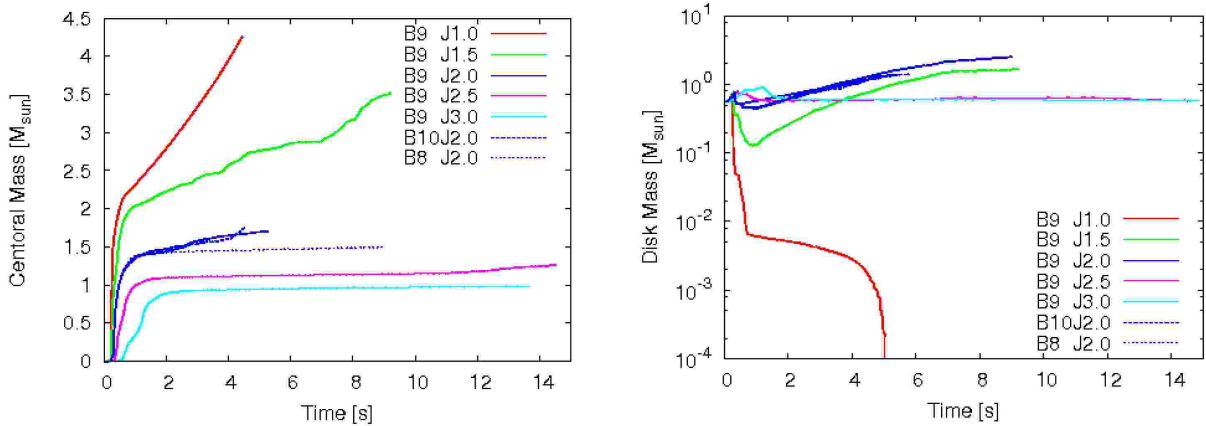}
\caption{Time evolutions of the masses accreting to the central objects (left) 
 and the masses of the accretion disks (right). 
Top 5 lines of red, green, blue, purple, and water correspond to 
the models with the same initial magnetic field of $10^{9}$ G but with
 $j = j_{\mathrm{lso}}, 1.5j_{\mathrm{lso}}, 
2 j_{\mathrm{lso}}, 2.5 j_{\mathrm{lso}}$, and $3 j_{\mathrm{lso}}$, respectively. 
For the case of $j = 2 j_{\mathrm{lso}}$ (blue), the dashed, solid, and dotted line 
 show the variation of the initial magnetic fields $(B_0 = 10^{10}, 10^9$, and $10^8$ G), respectively. It can be shown that 
for rapidly rotating models, the mass accretion rate to the center becomes smaller 
(left) and the accretion disks become 
 heavier (right). To estimate the disk mass, we count the mass elements 
 which are nearly in the hydrodynamical equilibrium near the equatorial plane.} 
\label{time_Mass}
\end{figure} 

The left panel of Figure \ref{time_Mass} shows the time evolution of 
the central mass for some representative models, where the central mass
 is defined to be the baryonic mass accreted 
onto the central object through the inner boundary. 
It is shown that the central mass grows larger and more quickly for the 
 models with the smaller initial angular momentum. This is simply due to the smaller 
centrifugal forces. 
  In fact, the slowest rotation model of B9J1.0 shows a fastest increase, 
and as a result, the mass of the accretion disk becomes lightest (right panel). 
 Such a feature seems close to the so-called dwarf disk, 
in which matter is swallowed to the center with nearly a 
free-fall velocity in the equatorial plane
(\citet{lee06,lopez09}). 
Here it should be noted that very rapid growth of the
 central mass has been predicted to affect the collapsar ability to produce jets
 \citep{jani08a,jani08b},
 which we observe actually in our simulations and will be explained
 in the following section. 
  For all the models with J1.0, on the other hand, 
we do not observe any MHD jets owing to inefficient 
 winding of the magnetic fields. Moreover neutrino luminosities from very
 thin accretion disks for the models is typically 
 about 2 orders-of-magnitudes smaller than for the other models. 
So, we regard such models as inadequate for the progenitor of 
collapsars and focus on the models with 
$j \ge 1.5 j_{\mathrm{lso}}$ in the subsequent sections. 
 Here, we define the critical angular momentum to be the minimum angular momentum to form the stable and thick disk. In our case, the critical value is 
$j = 1.5 j_{\mathrm{lso}}$. 

 From the left panel, it can be also seen that the increase of the central mass becomes
  slightly larger for models with larger initial magnetic fields (compare blue lines).
 This is likely due to the angular momentum transport via the magnetic fields, which 
 enhances the matter accretion to the center. 
 However the difference of the initial angular momentum is more 
 decisive to capture the main dynamical features among the computed models, 
 on which we focus in the following.

\subsubsection{MHD outflows and neutrino luminosity}
\label{subsec:lumi_outflow}
To extract the general features furthermore among the models, we focus on 
the properties of MHD outflows and neutrino luminosities.
It is noted that both of them is helpful to understand the energy sources 
 for powering the GRBs namely via magnetic and/or neutrino-heating 
mechanisms.  
 More specifically speaking, the information of the neutrino luminosity is indispensable
  to estimate the neutrino energy deposition from the accretion disk.
 And the formations of MHD outflows is also important, 
because it can evacuate the funnel for the secondary 
 jets along the rotational axis, which could be the birthplace of relativistic fireballs.

\begin{table}[tb]
\begin{center}
\begin{tabular}[c]{|c|c|c|c|c|}
\hline
 Model Name & J1.5& J2.0& J2.5& J3.0\\
\hline
B10& 
\shortstack{ $\bigcirc$ (TYPE II)  \\ $1.1 \times 10^{52}$ erg/s \\ 3.4 s}& 
\shortstack{ $\bigcirc$ (TYPE II) \\ $4.5 \times 10^{51}$ erg/s \\ 5.8 s}& 
\shortstack{ $\bigcirc$ (TYPE I) \\ $1.6 \times 10^{50}$ erg/s\\ 7.7 s}& 
\shortstack{ $\bigcirc$ (TYPE I) \\ $9.0 \times 10^{49}$ erg/s \\ 9.3 s}
\\ \hline
B9& 
\shortstack{ $\bigcirc$ (TYPE I) \\ $1.6 \times 10^{52}$ erg/s \\ 9.2 s}& 
\shortstack{ $\bigcirc$ (TYPE I) \\  $5.1 \times 10^{51}$ erg/s\\ 5.3 s}& 
\shortstack{ $\times$\\  $1.4 \times 10^{50}$ erg/s \\ 12 s}& 
\shortstack{ $\times$ \\ $2.5 \times 10^{49}$ erg/s \\ 14 s}
\\ \hline
B8& 
\shortstack{ $\times$  \\ $1.8 \times 10^{52}$ erg/s \\ 4.3 s}& 
\shortstack{ $\times$  \\ $8.5 \times 10^{51}$ erg/s \\ 6.0 s}& 
\shortstack{ $\times$  \\ $1.7 \times 10^{50}$ erg/s \\ 10 s}& 
\shortstack{ $\times$  \\ $4.5 \times 10^{49}$ erg/s  \\ 12 s}
\\ \hline
\end{tabular}
\caption{Properties of MHD jets and neutrino luminosities.
Contents of each cell is, whether the MHD jets are formed
(yes or no indicated by $\bigcirc$ or $\times$) with the different formation 
 mechanisms indicated by TYPE I or TYPE II (top),  the neutrino luminosity (middle)
 estimated at the epoch (bottom) when the accretion disks become almost stationary. 
While the success or failure of the jet-launch depends 
  both on the initial strength of magnetic fields and rotation, 
   the neutrino luminosity is shown to be predominantly determined by the initial 
 rotation rates.}\label{Table:lumi_outflow}
\end{center}
\end{table}

Top column of each cell of Table \ref{Table:lumi_outflow} 
indicates whether the MHD jets are formed 
(yes or no indicated by $\bigcirc$ or $\times$).
TYPE I or II indicates the difference of the 
formation process of the MHD jets, which we will explain in detail 
 from the next section (section \ref{jets}). 

The quantities of the middle column show the neutrino luminosity 
(sum of all the neutrino species $\nu_e$, $\bar{\nu}_e$, and $\nu_{X}$) 
estimated at the epoch when the accretion disks becomes 
 stationary(bottom) (e.g., typically  $\sim$ 4 sec in  Figure \ref{time_Mass}).
 We find that the neutrino luminosities become higher for slower rotation models.
 This is because the accretion disks can attain higher temperatures due to 
the gravitational compression. It is interesting to note that the luminosities
  tend to become smaller for strongly magnetized models with relatively 
 smaller angular momentum (J $\leq$ J2.5). This is mainly 
 because the gravitational compression is hindered by the magnetic forces confined 
 in the disks. More detailed analysis of the luminosities is given in 
 section \ref{subsec:lumi}.

\subsection{Formation of Magnetically-dominated  Jets}\label{jets}
In Table 2, we categorized the launching of the MHD jets in two ways.
Here it should be noted 
that we distinguish the collimated outflow as `` jets '' 
 where their half-opening angle is less than $10^\circ$. 
Before discussing the details of each type 
in section \ref{subsec:outflow}, 
we mention the amplification of the toroidal magnetic fields and the subsequent 
formation of the magnetic outflows from the accretion disks, which 
 precedes the jet formations.

\subsubsection{Amplification of magnetic fields and the outflows from accretion disks}\label{subsec:magamp}

\begin{figure}[tbd]
\begin{center}
\epsscale{1.0}\plotone{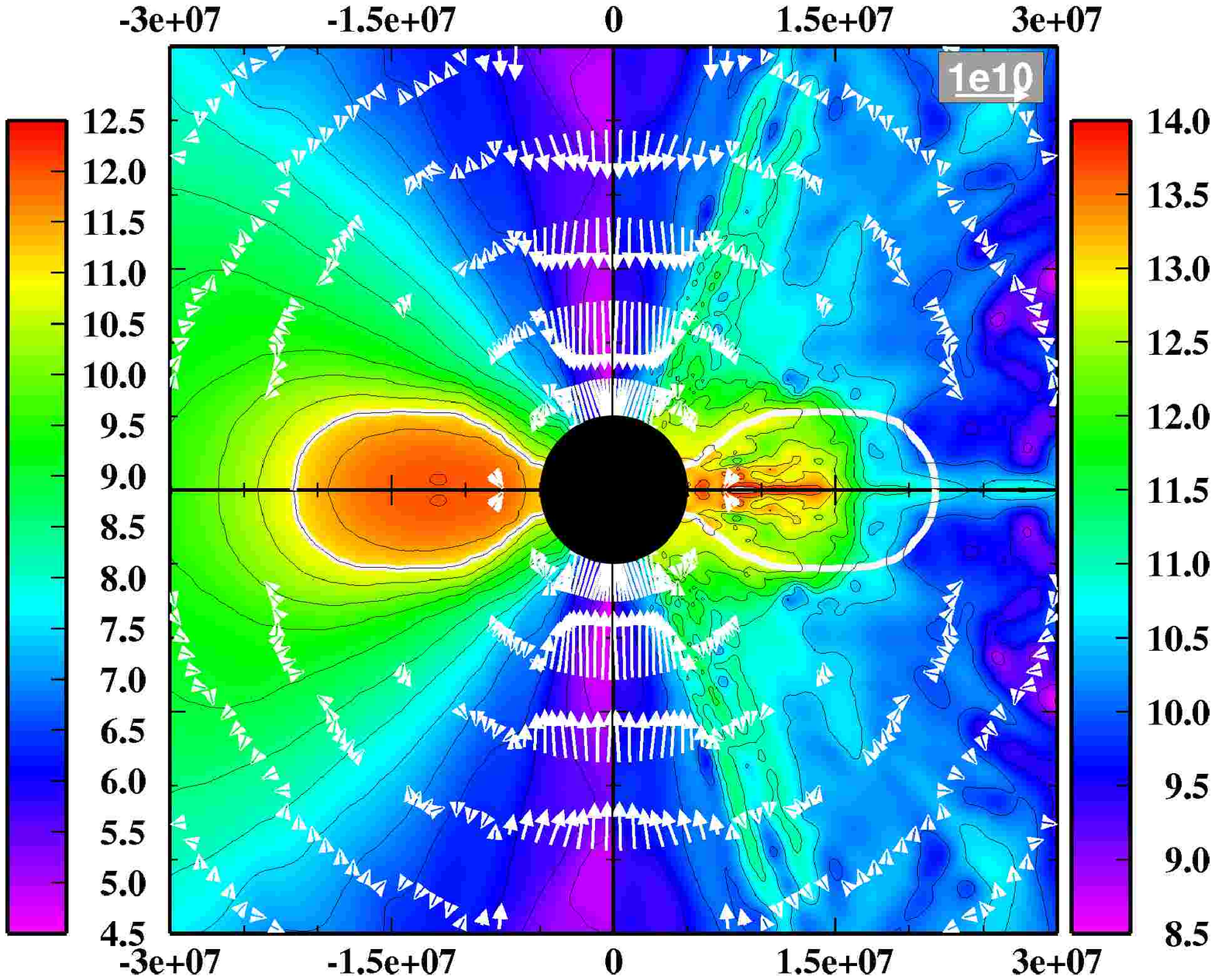}
\caption{Logarithmic contour of 
 density (left) and poloidal magnetic field (right) 
for B9J1.5 model, 
just before launching an outflow at 7.99 s. 
The white solid line (right) denotes the area 
where the density is equal to $10^{11}\ \gpcmc$, showing 
 the surface of the accretion disk.
The strong poloidal fields inside the accretion disk in the vicinity of 
the equatorial plane (right) coincide with the high density regions (left).
 This means that the poloidal fields are amplified mainly by compression. 
Note that the velocity fields are drawn by the white arrows, and 
the length is normalized by the scale shown in top right edge of the box 
(here, $10^{10} {\rm cm}~{\rm s}^{-1}$).
The central black circle (50 km in radius) represents the inner boundary 
of our computations. 
In the following figures, density line, velocity field, 
and the inner boundary are depicted in the same manner. }
\label{B9J1.5dBp}
\end{center}
\end{figure}
 
Since the initial models investigated here are assumed to have only the poloidal
fields (section \ref{sec:inimodel}), 
 the key ingredients for amplifying the toroidal fields 
are the compression of the poloidal fields 
and the efficient wrapping 
of them via differential rotations. 
In addition, MRI should also play an important role, 
whose wavelength of the fastest growing mode is 
given by 
$\lambda \sim 5 
\Bigl( \frac{300\psec}{\Omega}	  \Bigl) 
\Bigl( \frac{B}{10^{12}\gauss}	  \Bigl) 
\Bigl( \frac{10^{10}\gpcmc}{\rho} \Bigl)^{1/2} \rm{km}$ 
\citep{balb98}. 
Here, putting the typical physical values of the disk, 
our numerical grids are insufficient to capture MRI at 
 earlier phase when the magnetic field is weak, but can handle it 
in the later phase when the magnetic field gets stronger.
In this sense, the discussion below, which fails to include the effects of 
 MRI completely, should give a lower bound for the field amplification.
 Discussions about MRI are given with numerical tests in section \ref{tests}. 
 
Figure \ref{B9J1.5dBp} shows 
the distributions of density and poloidal fields of model B9J1.5
 at 7.99 s, just before the launch of the MHD outflows from the accretion disk.
The density takes its maximum value at around 120 km in the equatorial plane
 (left panel), in which the poloidal fields are strong because the higher 
compression is achieved (right panel). 
The white solid line in the right panel indicates
 the surface positions of the accretion disk. So the amplifications of the 
poloidal fields occur most efficiently inside the accretion disk.
It is noted here that the disk is gravitationally stable because 
the adiabatic index ($\gamma$) inside the disk becomes greater than $4/3$ due 
to the contribution of the non-relativistic nucleon ($\gamma = 5/3$) 
photodissociated from the iron nuclei. 

\begin{figure}[tb]
\begin{center}
\epsscale{1.0}\plotone{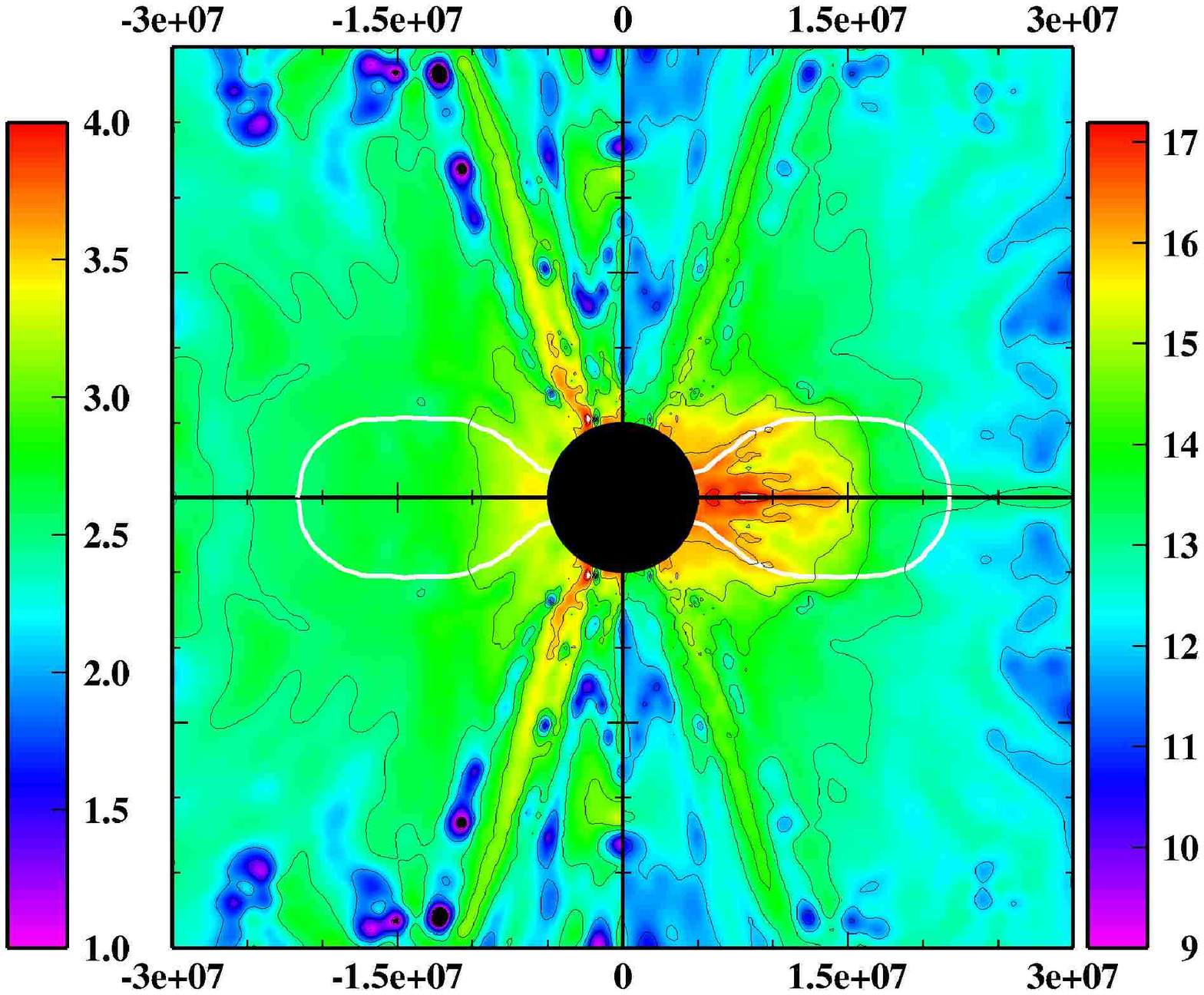}
\caption{Same as Figure 2 but for logarithmic contour of 
 differential rotation (:$Xd\Omega/dX$) (left) and amplification rate 
of toroidal magnetic field (:$d B_{\phi}/dt$) (right).
Comparing with the right side of Figure \ref{B9J1.5dBp}, 
 the amplification rates become larger where the poloidal fields and 
 the degree of the differential rotation are stronger.}
\label{B9J1.5domegadBphi}
\end{center}
\end{figure}

As for the toroidal fields, Figure \ref{B9J1.5domegadBphi} shows that their
 amplification rates are highest also inside the disk (right), because the 
 degree of the differential rotation is large there (left). 
In previous collapsar simulations assuming much larger angular momentum initially
(e.g., \citet{prog03c,fuji06}), 
it seems to be widely agreed that the differential rotation is a primary agent to 
 amplify the toroidal fields. 
On the other hand, our results show that for 
long-term evolution of relatively slow rotation models 
(see also \citet{prog05}), 
the amplification of the poloidal fields by compression 
 is preconditioned for the amplification of the toroidal fields. 

\begin{figure}[tb]
\begin{center}
\epsscale{1.0}\plotone{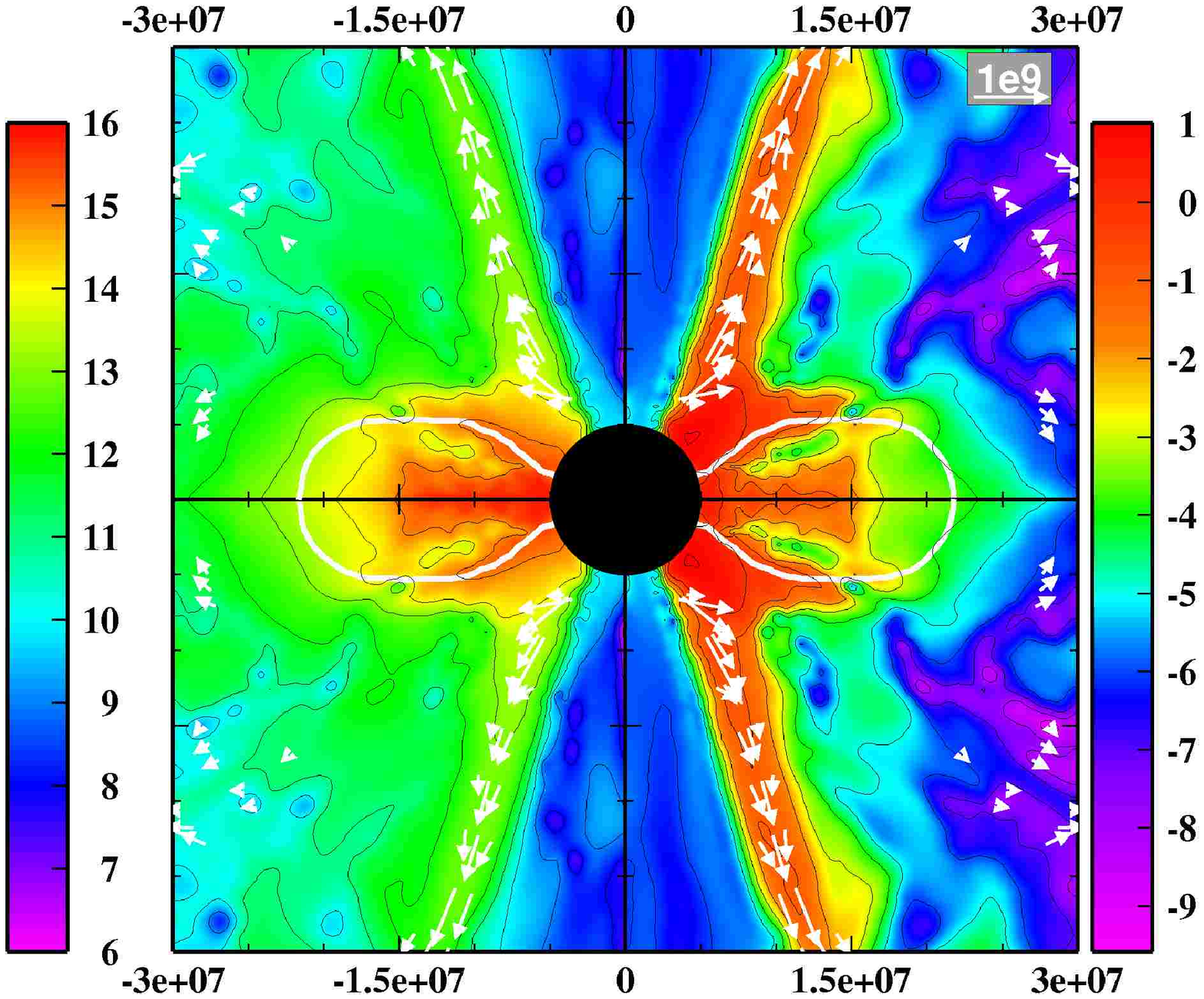}
\caption{Same as Figure \ref{B9J1.5domegadBphi}, but for 
 the logarithmic contour of toroidal magnetic fields (left) and 
the inverse of plasma beta (e.g., $\beta_{\rm mag}^{-1} \equiv 
\bold{B}\cdot\bold{B}/8\pi/p$) (right).
It is shown that the toroidal fields are amplified up to $10^{16}\ \gauss$ (left), 
and that the MHD outflows are driven from the accretion disk (right). 
In this figure, we plot the velocity fields only for
$V_r > 10^8 \cmps$ for illuminating the outflows, where $V_r$ is the radial velocity.} 
\label{B9J1.5Bphipmag}
\end{center}
\end{figure}

The left panel of Figure \ref{B9J1.5Bphipmag} shows that the
toroidal fields are amplified as high as $10^{16}$ G inside the disk, 
where the amplification 
rate is indeed high (see Figure \ref{B9J1.5domegadBphi} (right)). 
The ratio of the magnetic 
 to rotational energy at this moment is about $\sim 10 \%$, which is close  
 to the saturation level of the field growth as shown in \citet{shib06}.  
 From the right panel, 
 it can be seen that the magneto-driven outflows, in which the magnetic pressure 
dominates over the matter pressure, are produced in the 
vicinity of the accretion disk. Looking carefully, the amplification rates are 
 higher near along the equator (Figure 3 (right)) and decrease as the distance 
to the equator gets larger vertically (perpendicular to the equator).
We find that the vertical gradient of the magnetic pressure near the surface of the disk 
 can drive the MHD outflows. It is interesting that the propagation of the outflows 
is not along the rotational axis, but slightly off-axis. This is because the  
ram pressures just along the rotational axis, free from the 
centrifugal forces, are highest, and thus the magnetic pressures cannot overwhelm
 the ram pressure there.
Following these magnetic outflows from the disk, 
 the MHD jets are formed in the two ways (namely type I or II), 
which we will explain from the next section.

\clearpage

\subsubsection{Two types of MHD jet formation}\label{subsec:outflow}

When the magnetically-dominated outflows mentioned above 
are so strong enough to come out of the central iron cores without shock-stall, we 
call them as type II jets. Even if these prompt outflows stall at first, 
we find an another way of launching jets (type I), whose formation processes 
are a little bit complicated than for type II. We first explain type II 
in the following.


\begin{figure}[tbd]
\begin{center}
\epsscale{1.00}\plotone{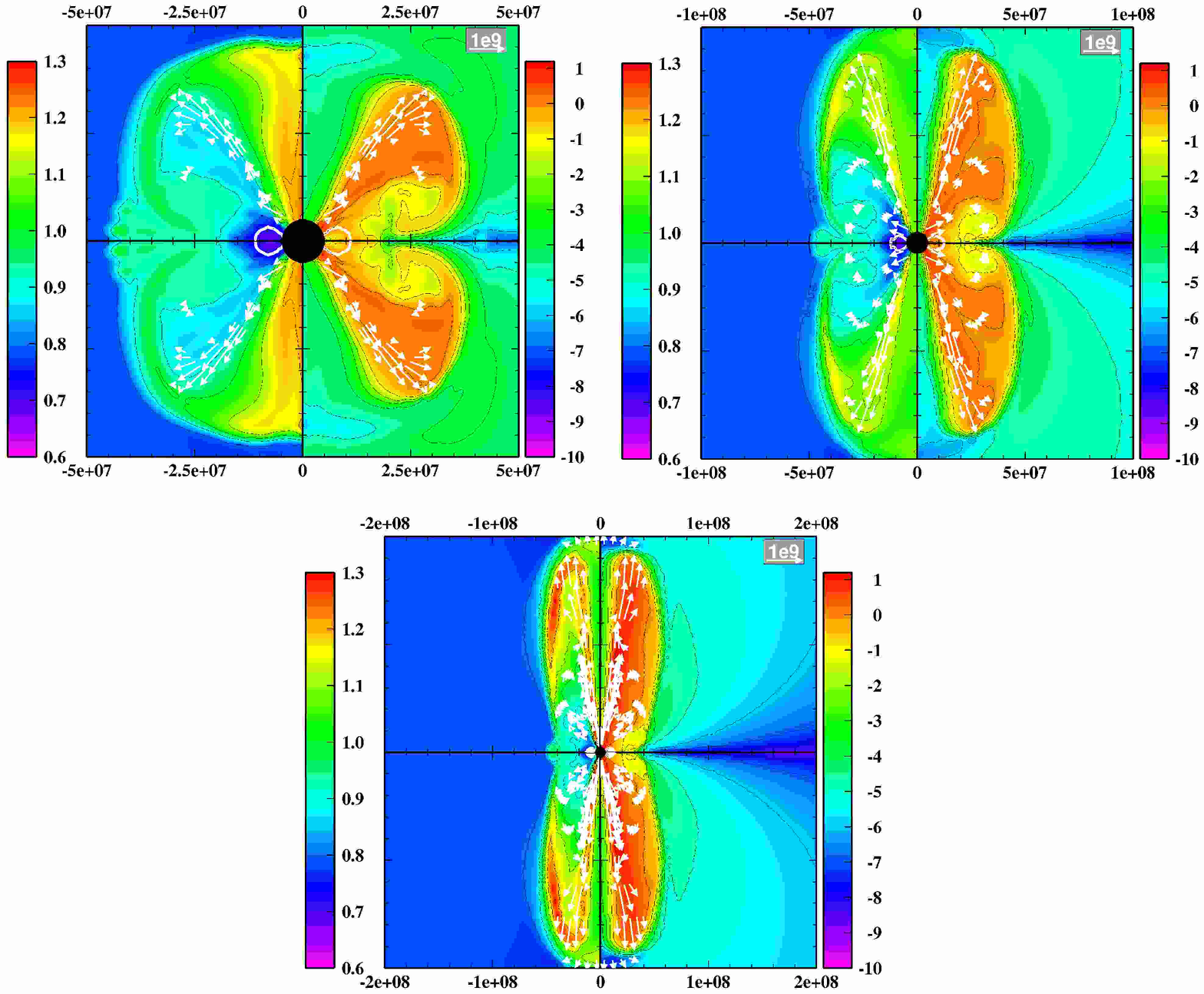}
\caption{Evolution of MHD jets launched from the accretion disk (Type II jets).
In each panel, logarithmic contour of 
 entropy (left side) and the inverse of plasma beta 
(e.g., $\beta_{\rm mag}^{-1} \equiv 
\bold{B}\cdot\bold{B}/8\pi/p$) (right side) are shown at 1.43 s (top left), 
1.87 s (top right), and 2.11 s (bottom) for model B10J1.5.
Note the difference of the length scales among panels. 
Without shock-stall, magnetically-driven outflows ($\beta_{\rm mag}^{-1} \gtrsim 1$) 
come out of the iron core ($\sim 3000$ km in radius).}
\label{B10J1.5spmag}
\end{center}
\end{figure}

\begin{figure}[tbd]
\begin{center}
\epsscale{1.0}\plotone{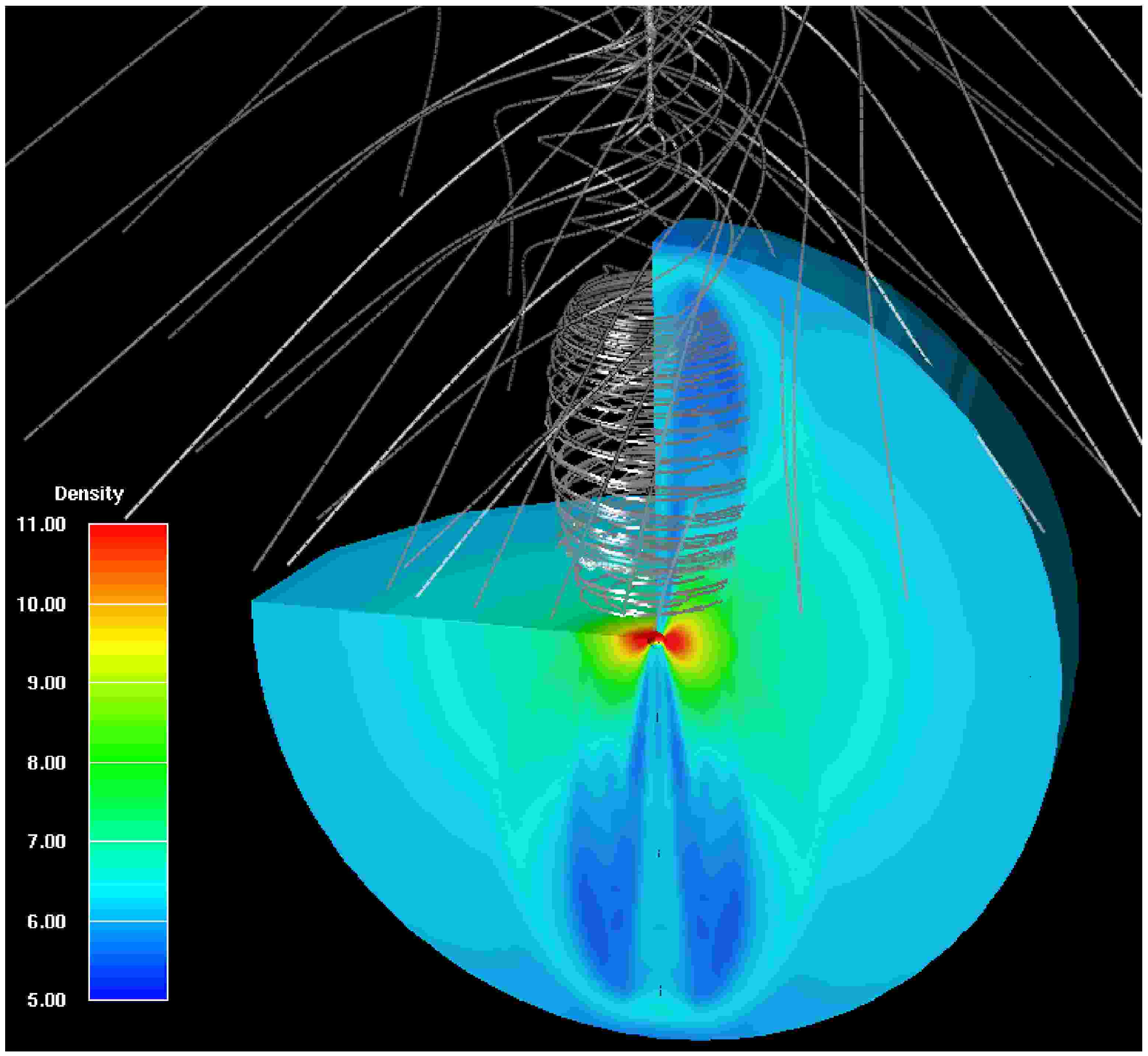}
\caption{Three dimensional plot of density with the magnetic field lines
 (silver line) for B10J1.5 model near at the moment of the shock break-out (2.11 s). 
Color contour on the two-dimensional slice represents the logarithmic density. 
The outer edge of the sphere colored by blue represents the radius of 
$2 \times 10^8 \cm$. The outflow is shown to be driven by the so-called magnetic tower, 
i.e., by the toroidal fields tangled around the rotational axis. 
Note that the 
field lines outside the bluerish region, seen to be more weakly twisted than inside, 
come mainly from the preshock region.}
\label{B10J1.5dB_3D}
\end{center}
\end{figure}

Figure \ref{B10J1.5spmag} shows the evolution of the MHD outflow for model B10J1.5,
 from its initiation near from the inner edge of the accretion disk (top left),
 propagation along the polar axis (top right), 
till they come out of the iron core (bottom). Note the difference of the length scales 
 in each panel. Among the computed models, this model has the strongest 
initial magnetic fields with smallest angular momentum (e.g., Table 2). 
 The toroidal fields can be much stronger than for other models, by the 
 enhanced compression of the poloidal fields inside the disk, as mentioned 
  in \S \ref{subsec:magamp}. 
As a result, the MHD outflow is so strong that they do not 
 stall, once they are launched. In fact, the outflow is shown to be kept 
magnetically-dominated (inverse of the plasma beta greater than 1, right-hand side 
 in Figure \ref{B10J1.5spmag}) till the shock break-out.
 As the disk becomes more compressed, the magnetic pressure of toroidal fields 
 and its vertical gradient inside the accretion disk become larger, which acts to 
  push the outflow vertically more strongly. In the early phase of 
the jets (top left), the sideway expansion of the outflow 
 is suppressed by the external pressure $P_{\mathrm{ext}}$,
 which is determined by the ram pressure of a freely falling fluid with 
velocity of $v_{r_{\rm ext}}$ as $P_{\mathrm{ext}} \approx \rho v_{r_{\rm ext}^2} \approx \rho GM/r_{\rm ext}$, with $r_{\rm ext}$ being the width of the outflow.
 $r_{\rm ext}$ is 
determined by the balance equation of $B^2/8\pi = P_{\mathrm{ext}}$, which gives a reasonable value. This confinement promotes the outflow to keep progressing 
 vertically. As the outflow propagates rather further from the center (top right),
 the outflow begins to be collimated due to the magnetic hoop stresses, 
 and keep their shape till the shock-breakout (Figure \ref{B10J1.5dB_3D}).
It is interesting to notice that 
the ram pressure just along the polar axis is so 
 large that no outflow is formed there, and that there stays 
 a polar funnel, 
 where the material accretes onto the central object. 
 We speculate that the formation of the funnel in such an early phase could 
 possibly provide nice environments as a birthplace of fireballs, because 
the neutrino heating from the disk could be sufficiently high at the epoch 
as will be discussed in section 5.2.

It should be noted here that the above outflow driven by the toroidal 
fields is essentially same as the 'magnetic tower' which was first introduced 
by \citet{lynden96,lynden03} in the context of active galactic nucleus, and applied to 
the collapsar environments \citep{uzde07a,uzde07b}.
However in the analytic models by \citet{uzde07a,uzde07b}, the driven mechanism of 
 the tower are assumed to be the winding of the magnetic fields threaded 
 in the planar accretion disks with no vertical structures inside. Our simulations 
 suggest that the origin comes from the vertical gradient 
of the twisted toroidal fields inside the accretion disk. 
\clearpage

\begin{figure}[tbd]
\begin{center}
\epsscale{1.0}\plotone{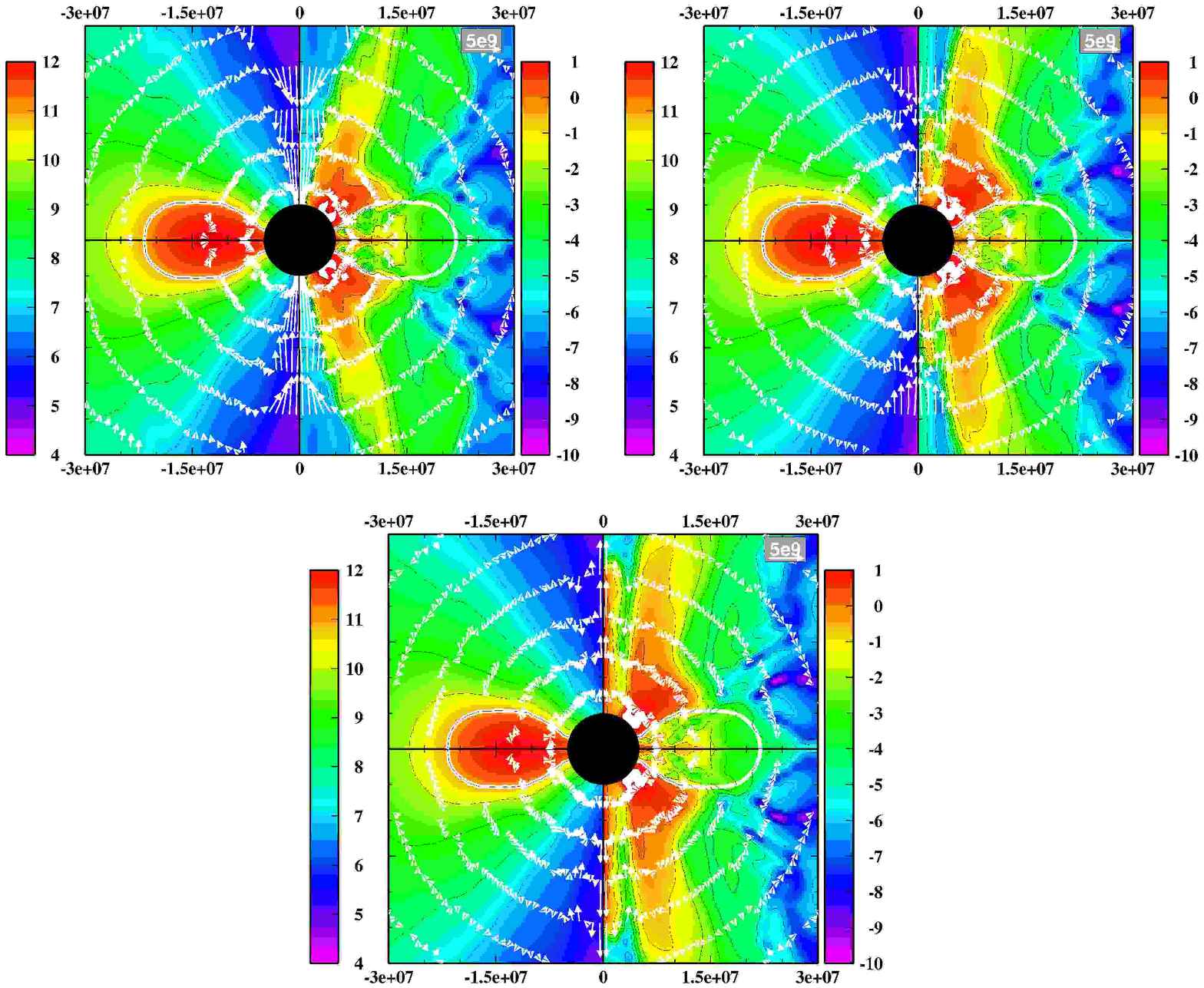}
\caption{Same as Figure 4 but for model B9J1.5
 at 8.115 s (top left), 8.148 s (top right), and 8.154 s (bottom), 
 showing the moment of the formations of jets in type I.
In the top right panel, very narrow magneto-driven regions along the rotational axis 
are produced by inflows of the 
accreting material in the equator to the polar regions.
 Such flow-in materials are shown to start propagating 
along the rotational axis (bottom).}
\label{B9J1.5dpmag}
\end{center}
\end{figure}

\begin{figure}[tbd]
\begin{center}
\epsscale{1.0}\plotone{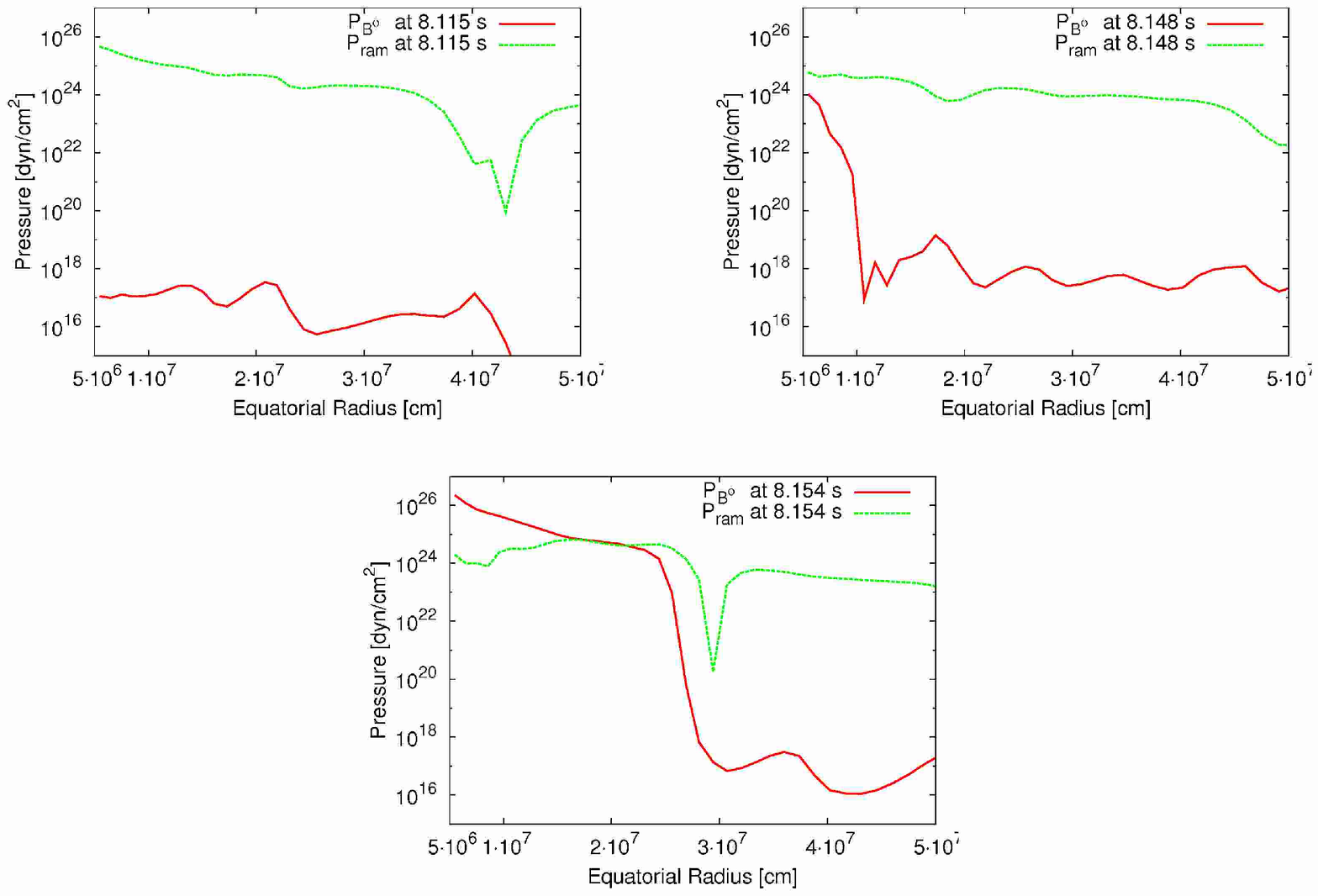}
\caption{The magnetic pressure of toroidal fields 
 vs. the ram pressure for model B9J1.5 at 8.115 s (top left), 
8.148 s (top right), and 8.154 s (bottom) along the rotational axis, respectively.
 Note that the timescales correspond to the ones
 in Figure \ref{B9J1.5dpmag}.
Only after the magnetic pressure becomes dominant over the ram pressure (bottom), 
the materials inside $\sim$ 200 km begin to propagate outwards.}
 \label{B9J1.5press}
\end{center}
\end{figure}

Now we move on to discuss the jets in type I, by taking model B9J1.5 as 
an example. In this case, the magnetic outflow from the accretion disk 
 is not as strong as model B10J1.5, and stalls at first in the iron core 
(see butterfly-like regions colored by red in the top left panel (right-hand side) 
 in Figure \ref{B9J1.5dpmag}). In the top right panel, 
 very narrow regions near along the rotational axis are seen to be produced 
 in which the magnetic pressure dominate over the matter pressure 
 (colored by red in the right-hand side). Such regions are formed by turbulent inflows 
of the accreting material from the equator, crossing the butterfly-like regions
 outside the disk, to the polar regions. 
Such flow-in materials obtain sufficient 
 magnetic amplifications when they approach to the rotational axis where
 the differential rotation is stronger, leading to the formations of the 
MHD outflows along the rotational axis (bottom).

\begin{figure}[tb]
\begin{center}
\epsscale{1.0}\plotone{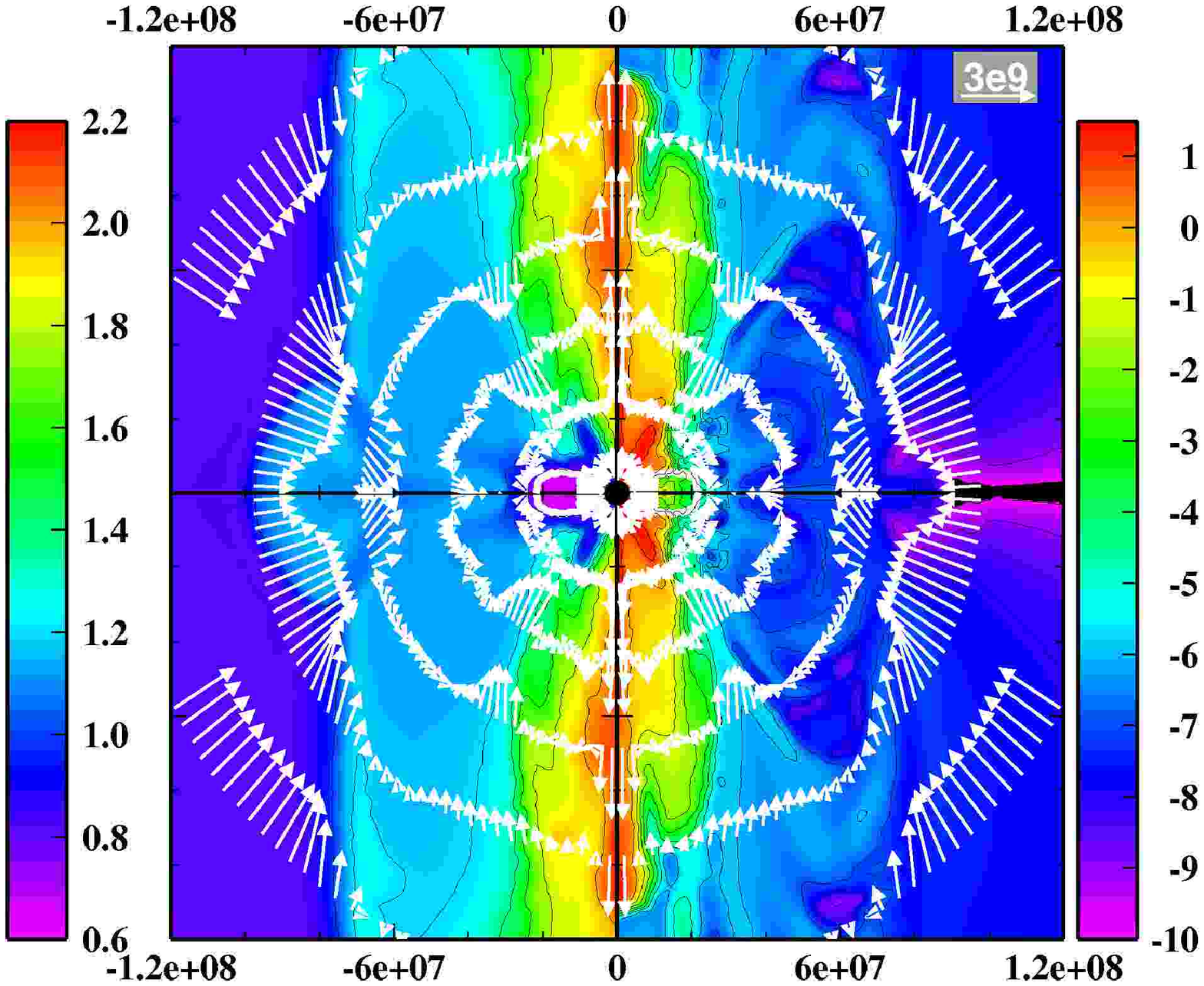}
\caption{Logarithmic contour of 
 entropy (left) and  inverse of plasma beta (right) 
for B9J1.5 model at 8.23 s. 
 The narrow magneto-driven explosions are shown near the rotational
 axis, which is produced by the type I mechanism (see text). 
 High entropy region ($\sim$ 20) outside the collimated jets (colored by light-blue 
 (left)) is a cocoon, which is produced by a fallback of the matter from 
the shock front.}
\label{B9J1.5spmag}
\end{center}
\end{figure}

Figure \ref{B9J1.5press} shows the magnetic pressure of the toroidal fields 
 and the ram pressure of the accreting material 
near along the rotational axis for the same time epoch 
 as in Figure \ref{B9J1.5dpmag}. The mentioned turbulent inflows to the 
polar regions happens at 8.148 s (top right), showing the increase of the magnetic pressure in the 
 central region within $100$ km. When the toroidal fields are amplified 
 enough to overwhelm the local ram pressure, $\rho v_r^2$ with $v_r$ being the radial 
 velocity (bottom), the flow-in materials begin to move outwards (bottom in Figure 
\ref{B9J1.5dpmag}). In fact, at about 70 ms later, the initiated outflows can 
 produce the well-collimated magneto-driven explosions, reaching $\sim 1000$ km 
 along the polar axis (Figure \ref{B9J1.5spmag}). 
Although the shape of the narrow jets are much like the ones obtained 
in previous collapsar simulations with much rapid rotation and strong magnetic fields 
(e.g., \citet{fuji06}) 
 or with slower rotation (e.g., \citet{prog05,mosc09}), 
it should be noted that their formation processes
 are different. The obtained outflow here is peculiar as a consequence of the 
 long-term evolution of collapsars, 
 which is produced by the interplay between the decreasing ram pressure and 
  the magnetic twisting of the turbulent inflow in the vicinity of the polar region. 

Now let us return to Table \ref{Table:lumi_outflow} again. 
Among the models with outflows, models B10J1.5 and B10J2.0 make
 the type II jets, and the other make the type I jets.
As mentioned, the type II is obtained for models with stronger 
 magnetic fields with relatively smaller angular momentum. 
If each of these conditions were not 
satisfied, the MHD outflows would stall at one time but 
with the subsequent revival (type I) or stall forever ($\times$ in Table 2). 

\clearpage

\subsubsection{Properties of MHD jets}\label{subsec:jet}

\begin{table}[tb]
\begin{center}
\begin{tabular}[c]{cccccc}
\hline
Model& $t_{\mathrm{jet}}\ [s]$& $M_{\mathrm{jet}}\ [\Ms]$& $E_{\mathrm{jet}}\ [\erg]$& $\Gamma_{\mathrm{jet}}$& $\Gamma_{\mathrm{jet,mag}}$\\
\hline
B10J1.5& $2.66$ & $1.3\times10^{-3}$& $5.6\times10^{48}$& $1.0024 (0.07c)$& $1.043(0.28c)$\\
B10J2.0& $3.80$ & $5.1\times10^{-4}$& $5.5\times10^{47}$& $1.00063 (0.04c)$& $1.0087(0.13c)$\\
B10J2.5& $6.33$ & $9.3\times10^{-4}$& $6.5\times10^{47}$& $1.0021 (0.06c)$& $1.052(0.31c)$\\
B10J3.0& $8.89$ & $1.7\times10^{-4}$& $2.4\times10^{47}$& $1.0077 (0.04c)$& $1.031(0.24c)$\\
\hline
\end{tabular}
\caption{Properties of MHD Jets. $t_{\mathrm{jet}}$ is the time when the jets 
come out of the iron core. $M_{\mathrm{jet}}$ and $E_{\mathrm{jet}}$ is
 the mass and the explosion energy of jets, respectively. 
 $\Gamma_{\mathrm{jet}}$ and $\Gamma_{\mathrm{jet,mag}}$ represent 
 the Lorentz factors (and the velocity normalized by the speed of light) 
and their maximum values estimated by taking into account the 
 magnetic energies, respectively. Note that all of them are estimated at $t_{\rm jet}$. }\label{Table:jet}
\end{center}
\end{table}

Now we proceed to look more in detail to the properties of the MHD jets.
Here we focus on the models with B10, because all the model sequence is accompanied 
 with the jets either type I or II (Table 2).

Table \ref{Table:jet} shows the mass, explosion energy, and Lorentz factor
 of jets at the moment of shock break-out. Here it is noted that the 
 explosion energy is estimated for the regions
 where the local energy is positive and the radial velocity is positive, 
indicating that the matter is not bound by the gravity 
(see the definition of $e_{\mathrm{local}} > 0$ in \ref{ap:energy}). 
 As seen in the table, the jet of model B10J1.5 has the largest explosion energy with 
 largest baryon loading. This is because the jet is type II as mentioned. 
 Since the jet is launched rather earlier ($\sim$ 1.9 s) than for type I, 
there is much material near the rotational axis, which makes the baryon-load 
of jets larger for the model.
For type I jets, no systematic dependence of the initial angular momentum on the 
 masses and the energies is found. We think that this is because the formation 
of the type I jets occurs by turbulent inflows as already mentioned in the previous section.

The similarity between types I and II is that
 the jets are at most subrelativistic (0.07$c$ for model B10J1.5) with 
the explosion energy less than $10^{49} \erg$.   
To see the maxim of the Lorentz factor in our computations, we boldly 
 assume that the magnetic energy of the fluid is fully converted to the kinetic energy,
 having in mind the dissipative process such as magnetic-reconnection
 ($\Gamma_{\mathrm{jet,mag}}$ in 
 Table \ref{Table:jet}). 
Even in this case, the jets become only mildly relativistic.
While the ordinary GRBs require the highly relativistic ejecta,
 we speculate that these mildly relativistic ejecta may be favorable for X-ray flashes
\citep{sode06,ghis07}, which is a low energy analogue of the GRBs.
 The propagation of the MHD jets can expel the matter 
 along the polar axis. Such baryon-poor environments may be a favorable cite
   for producing the subsequent jets, which could attain high Lorentz factors
 pushed by the magnetic outflows and/or heated by 
 neutrinos. 
 From the next section, we study the properties of neutrino 
luminosities obtained in our simulations and discuss how the neutrino heating, 
 albeit not coupled to the hydrodynamics here for simplicity, could have impacts on the 
jet formations.
\clearpage

\subsection{Properties of neutrino luminosity}\label{subsec:lumi}
\begin{figure}[tbd]
\begin{center}
\epsscale{1.0}\plotone{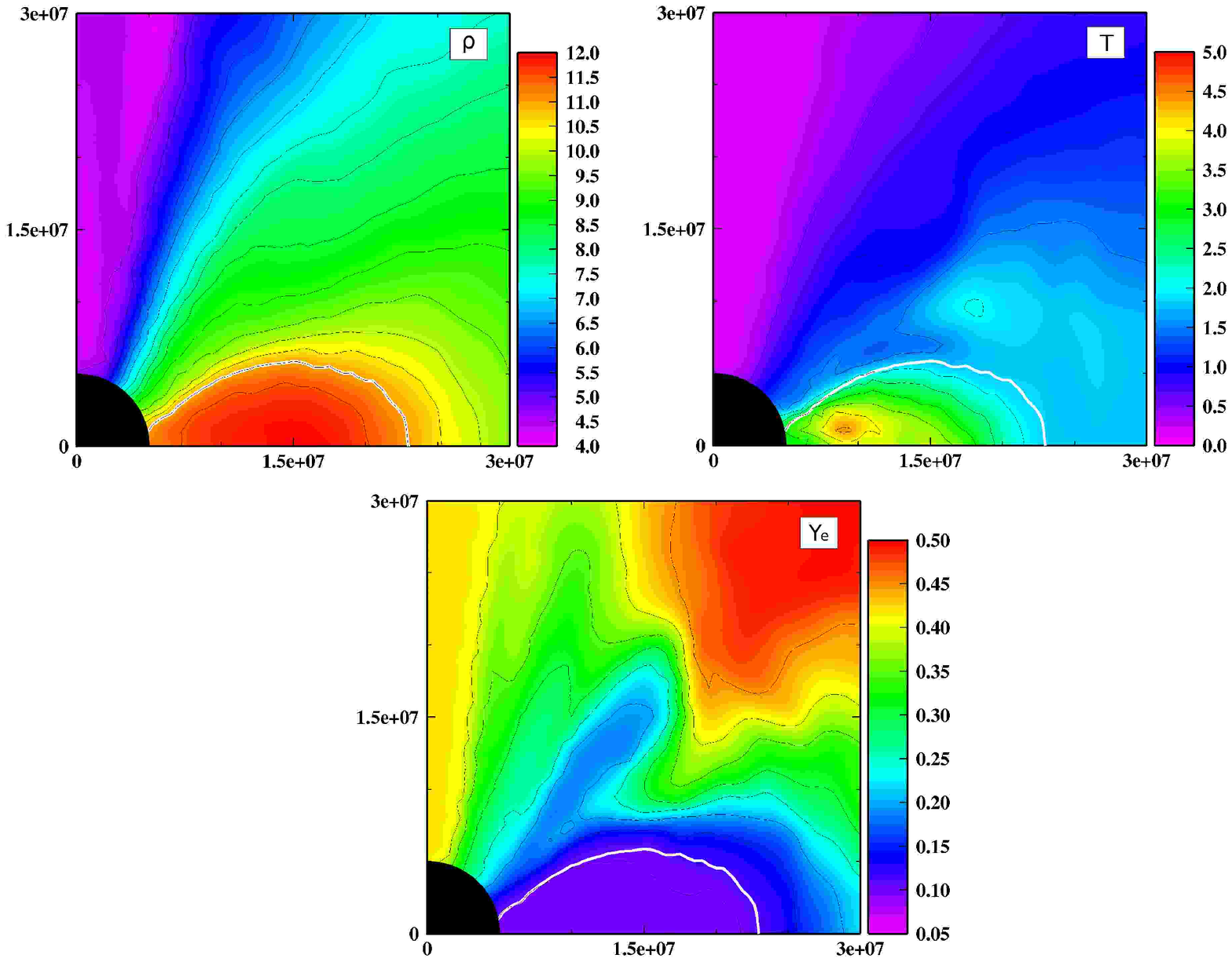}
\caption{Contour of 
 density [$\gpcmc$](top left), 
temperature [MeV](top right), 
and electron fraction (bottom) 
for model B9J1.5. These panels are for 9.22 s, when the accretion disk 
becomes stationary (see left panel of Figure 1). 
The white line marks $\rho=10^{11}\ \gpcmc$, representing the surface of the disk. 
Comparing with Figure \ref{B9J3.0dTYe} which has larger angular momentum initially, 
it can be seen that the disk here is more compact
 (top left) with higher temperatures (top right), and that the resulting enhancement of the  electron captures lowers $Y_e$ in the disk (bottom).}
\label{B9J1.5dTYe}
\end{center}
\end{figure}

As seen in Figure \ref{time_Mass} (left), 
 the disk masses for models 
with the MHD jets ($j \ge 1.5 j_{\mathrm{lso}}$), differ several times at most. 
On the other hand, the neutrino luminosities, which are 
 contributed mainly from the accretion disk, differ up to 
 3 orders-of-magnitudes (Table 2). This indicates that 
the local neutrino emissivities also 
change over several orders-of-magnitudes in the vicinity of the disk.
 In this section, by focusing on the structures of the disk,
 we discuss how the properties of the neutrino luminosities change 
for the models with different initial angular momentum.


\begin{figure}[tbd]
\begin{center}
\epsscale{1.0}\plotone{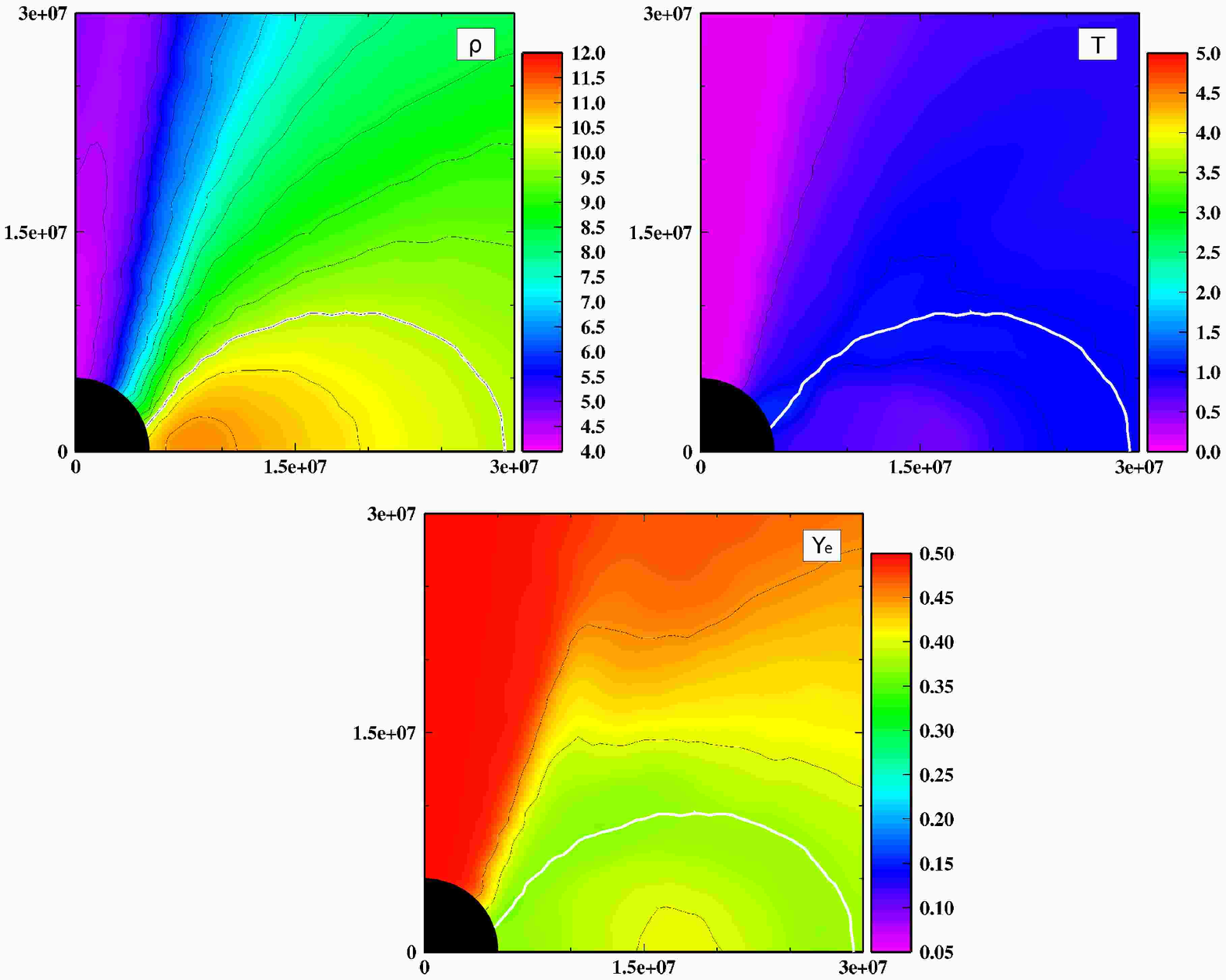}
\caption{Same as Figure \ref{B9J1.5dTYe} but for model B9J3.0.
Note here that the white line marks $\rho=10^{10}\ \gpcmc$. }
\label{B9J3.0dTYe}
\end{center}
\end{figure}

By comparing Figures \ref{B9J1.5dTYe} with \ref{B9J3.0dTYe}, 
 it can be seen that for smaller initial angular momentum (Figure 
 \ref{B9J1.5dTYe}), the disk becomes more compact
 (top left) with higher temperatures (top right), and that the resulting enhancement 
of the electron captures lower $Y_e$ in the disk (bottom).
 The enhanced compression is a primary reason for
the higher neutrino luminosities for slower rotating models in Table 2.

\begin{figure}[tb]
\begin{center}
\epsscale{1.0}\plotone{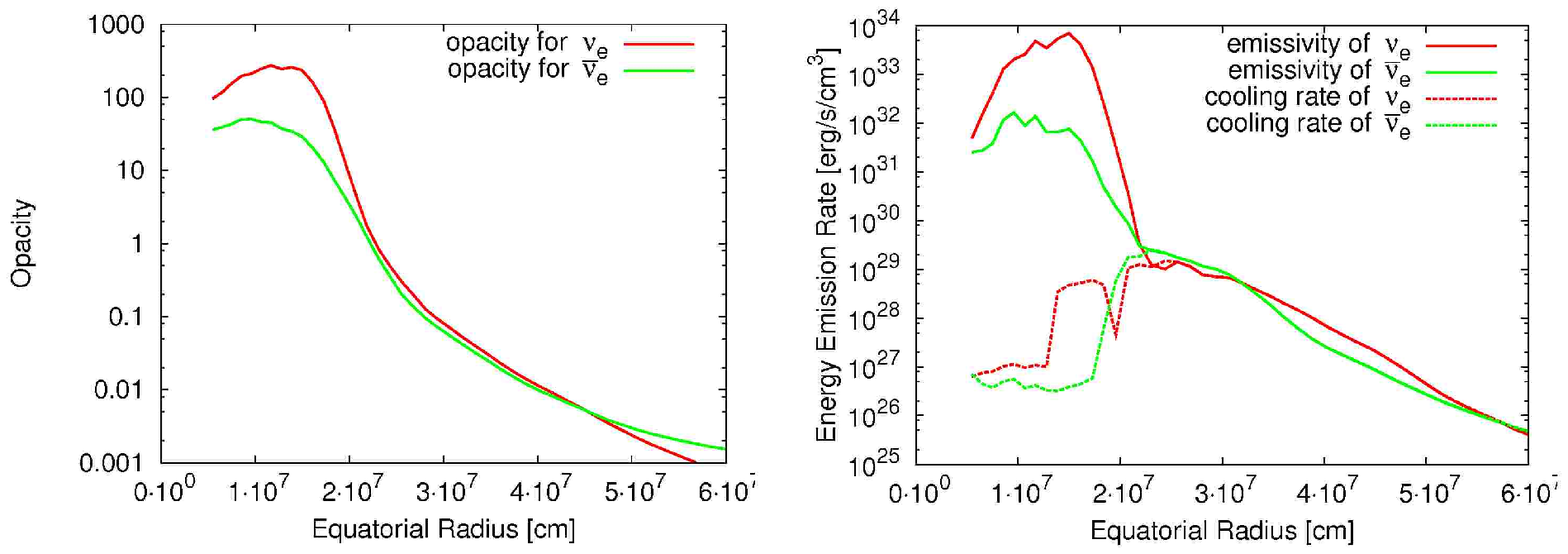}
\caption{Plots of opacity (left), emissivity (solid line, right), 
and cooling rate (dashed line, right) 
for $\nu_e$ (red) 
and ${\bar \nu}_e$(green) along the equatorial plane 
of model B9J1.5 at 9.22 s. 
The cooling rates are greatly reduced from the local neutrino emissivities 
 for the opaque regions ($\lesssim 200$ km).
For such regions, the higher suppression for $\nu_e$ makes its cooling rate almost 
comparable to that of $\bar{\nu}_e$.}
\label{B9J1.5opem}
\end{center}
\end{figure}

\begin{figure}[tbd]
\begin{center}
\epsscale{1.0}\plotone{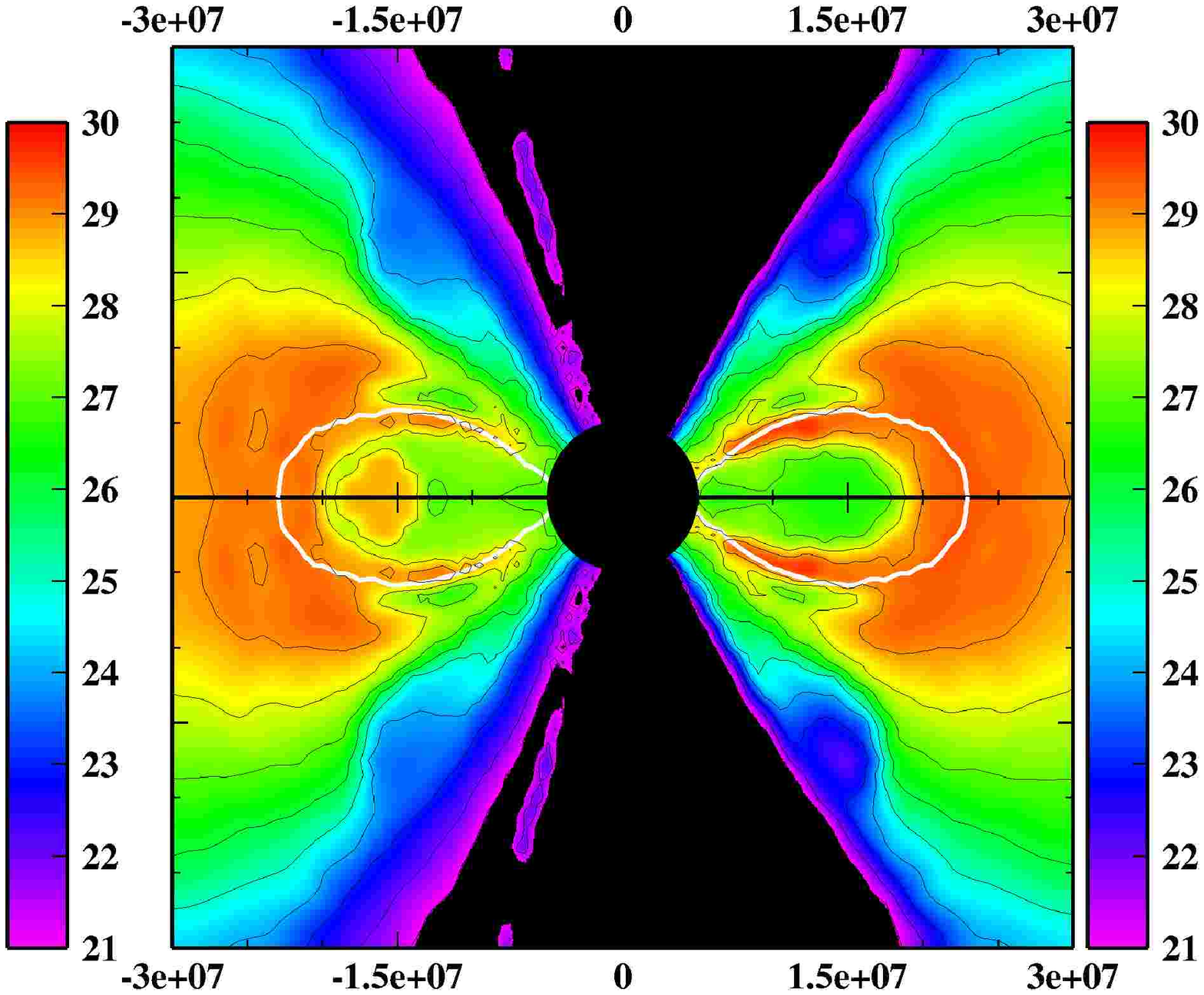}
\caption{Logarithmic contour of 
the cooling rates for $\nu_e$ (left), 
and for ${\bar \nu}_e$ (right) for model B9J1.5 model at 9.22 s. 
 It is shown that the luminosities of $\nu_e$ and ${\bar \nu}_e$ are almost
 comparable. Regions colored by black have lower values than the limit of the color 
legend, which is out of our interest here. The white line marks 
$\rho=10^{11}\ \gpcmc$, indicating the surface of the disk.}
\label{B9J1.5lnu}
\end{center}
\end{figure}

The left panel of Figure \ref{B9J1.5opem} shows that the disk, whose equatorial
 size is $\sim 200$km (top left in Figure 10), is very thick
 to neutrinos. In fact, the opacity inside the disk 
becomes up to 200 for $\nu_e$ inside the disk, which is one order-of-magnitude 
 larger than that for ${\bar \nu}_e$. 
The right panel shows that the local neutrino emissivities are highly suppressed 
 by $\exp(-\tau_{\nu})$ with $\tau_{\nu}$ being the opacity, which gives the 
 actual neutrino cooling rates in such a thick region (dashed lines). 
 The higher suppression for $\nu_e$ makes its cooling rate almost 
comparable to that of $\bar{\nu}_e$, which is also shown in Figure \ref{B9J1.5lnu}.
 It is noted that because of this high neutrino 
opacity inside the disk, the characteristic neutrino cooling timescale is typically 
more than four orders-of-magnitudes longer than the advection timescale.
 Thus the disk of our models is advection dominated flow.

\begin{figure}[tbd]
\begin{center}
\epsscale{1.0}\plotone{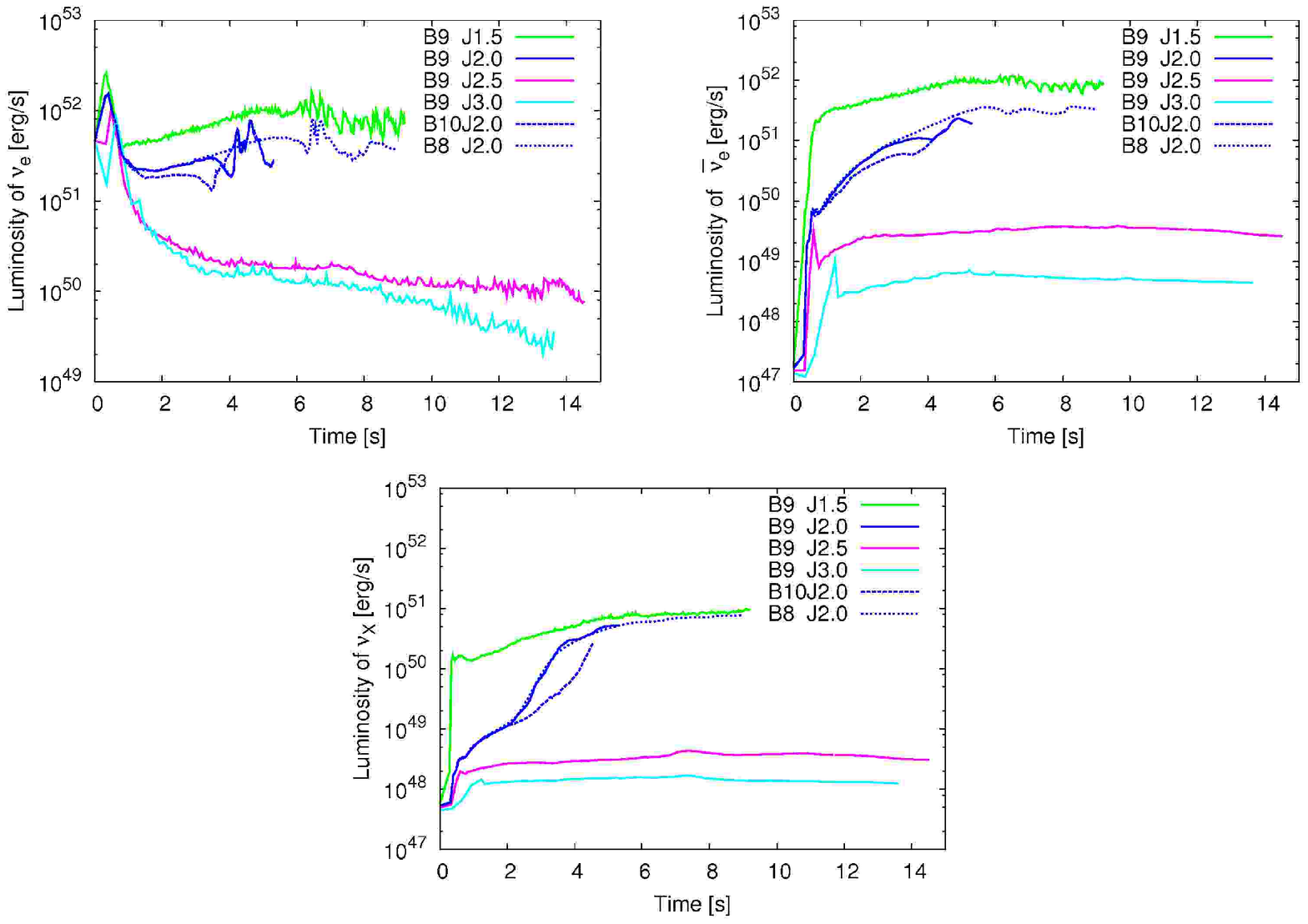}
\caption{Plots of time evolution of 
the luminosity for $\nu_e$ (top left), 
for ${\bar \nu}_e$ (top right),
and for $\nu_X$ (bottom). 
For slower models with J1.5 and J2.0, the neutrino luminosities of
 $\nu_e$ and ${\bar \nu}_e$ become almost comparable, which is 
 expected to be advantageous for energetizing the jets via neutrino heating.}
\label{lumi}
\end{center}
\end{figure}

Figure \ref{lumi} shows the evolution of neutrino luminosities for each neutrino
 species ($\nu_e$ (top left), $\bar{\nu}_e$ (top right), and $\nu_X$ (bottom)). 
 It is shown that for every species, the luminosities become larger 
for models with smaller angular momentum due to the mentioned higher compression,
 and also that the luminosities of $\nu_e$ and $\bar{\nu}_e$ are 
dominant over the ones of $\nu_X$.   
For models with J1.5 and J2.0, the luminosities of $\bar{\nu}_e$ 
become almost comparable to those of $\nu_e$, while the luminosities of $\bar{\nu}_e$
 are much smaller than  those of ${\nu}_e$  for more rapidly rotating models 
(compare top left and right panels in Figure 14 noting the difference of the vertical scales). On top of the effect of opacity mentioned above,
 this is because the higher compression leads to 
 the production of positrons more abundantly, which promotes the production 
 of $\bar{\nu}_e$ via the positron captures. 

Figure \ref{lumi} also depicts the effect of the magnetic fields on the luminosities
 (compare between B8J1.5, B9J1.5, and B10J1.5). For every neutrino species,
 the luminosities are shown to become smaller for models with larger initial magnetic
 fields. For stronger magnetized models, the disks expand more due to the 
 magnetic pressure inside, leading to the suppression of the contraction.

Among the computed models, the total neutrino luminosity becomes largest
$\sim 10^{52}$ erg/s for model with J1.5 series (e.g., Table 2).  
 It is well known that the energy deposition rate via pair neutrino annihilation
($\nu + {\bar \nu} \rightarrow e^{-}+ e^{+}$) is proportional to 
$L_{\nu}\cdot L_{{\bar \nu}}$ with being the neutrino and anti-neutrino luminosities,
 respectively \citep{salm99,asano1,asano2}. Thus almost the 
 equivalent luminosities of  $\nu_e$ and $\bar{\nu}_e$ is advantageous for 
 making the deposition rate larger for a given sum of the luminosities.
Due to the two factors, the models with smallest angular momentum 
computed here are expected to be most suitable 
for making the fireballs via neutrinos.

\clearpage

\section{Discussion}\label{sec:discussion}
Here we shall discuss limitations of our simulations and 
requirements towards more sophisticated numerical modeling of collapsar, such as 
 the effects of inner boundary (section \ref{subsec:jlso}), 
 the effects of neutrino heating (section \ref{subsec:heat}), and the 
importance of general relativity  (section \ref{subsec:kerr}). The latter two is
 neglected or treated by a simple approximation in this paper.  
Numerical tests are given in section  \ref{tests}.

\subsection{Effects of inner boundary} \label{subsec:jlso}

First of all, we discuss possible drawbacks due to 
the large inner boundary (50 km) taken in this simulation.  
For the Schwartzschild metric, the marginal stable orbit can exist for 
$j > 2 r_g c$ where $r_g$ is the gravitational radius. 
Such orbiting flow can lead to the formation of the shock on the equatorial plane, 
and is predicted to result in the formation of stable and thick disk 
as shown in \citet{prog03a,prog03b,lopez09}. 
To capture such a feature needs the inner boundary as small as 
$R_{\rm in} > 2 r_g$. However, our inner boundary is larger than the value. 
For example, models B10J1.5 and B9J1.5 have 
the central object with mass of $2.5 \Ms$ and $3.5 \Ms$ 
at the end of computation, which corresponds to 
$r_g = 7.5 {\rm km}, 10.5 {\rm km}$, respectively. Our inner boundary corresponds to 
$6.7 r_g, 4.8 r_g$ for each model. Therefore the critical angular momentum 
to form the stable accretion disk discussed in this paper could be 
 affected by the position of the inner boundary.  
Moreover the large inner boundary, which excises the inner edge of the accretion 
 disk, should lead to the underestimation of the
 neutrino luminosity and the resulting neutrino heating, which we will discuss in the 
 next section.
 To clarify 
 those points, we are now planning to implement  more compact inner boundary
 in the long-term evolution, which is computationally very demanding, 
thus as a sequel of this paper.

\subsection{Importance of Neutrino Heating} \label{subsec:heat}

\begin{figure}[tb]
\begin{center}
\epsscale{1.0}\plotone{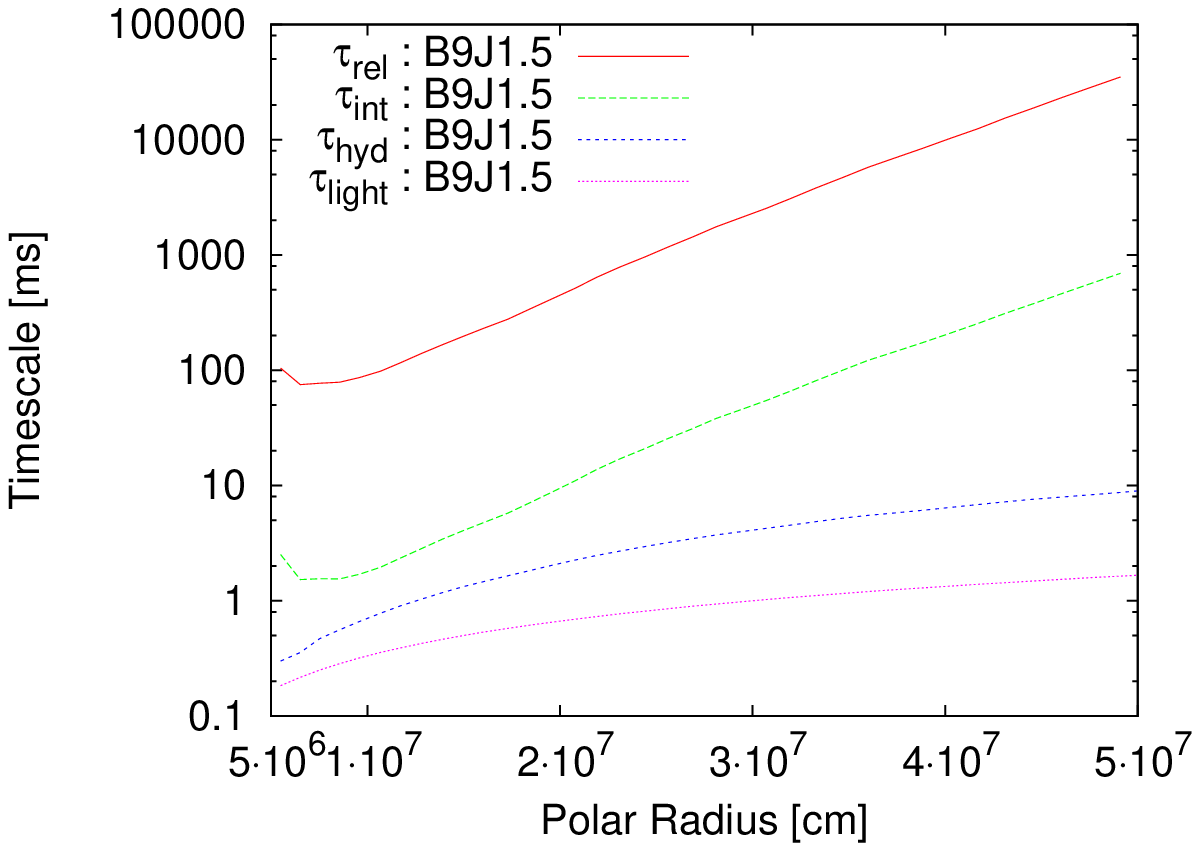}
\caption{Plots of 
 various timescale for matter ; the timescale to get relativistic 
by the neutrino heating $\tau_{\mathrm{rel}}$ (solid), 
the timescale to affect dynamics with neutrino heating 
$\tau_{\mathrm{int}}$ (dashed), 
the timescale of dynamical motion $\tau_{\mathrm{hyd}}$ (dotted), 
and the timescale of light crossing $\tau_{\mathrm{light}}$ (small dotted), 
at 9.22 s when the accretion disk reaches to the stationary state with the funnel 
 regions along the pole.}
\label{heat}
\end{center}
\end{figure}

We try to estimate the effects of the neutrino energy deposition via 
 neutrino pair annihilation ($\nu + {\bar \nu} \rightarrow e^{-}+ e^{+}$)
from the accretion disk to the polar funnel. The neutrino heating is important, however, not included in the simulations here for simplicity. 

In the following, we present an order-of-magnitude estimation of the heating rate.
To do so, we derive the heating rate with the special relativistic corrections 
 (see \ref{ap:srheat}). For simplicity, we take the so-called optically thin limit 
 in the accretion disk as in \citet{asano1} and consider the neutrino heating only along
 the rotatinal axis. It is noted that expect for the polar funnel, 
to make neutrino-heated outflows is hopeless due to the baryon contamination. 
By comparing timescales such as heating and advection,
we discuss how important the heating could be.  

Figure \ref{heat} shows various timescales.
  $\tau_{\mathrm{rel}} \equiv \rho c^2/q^+$ characterises the timescale
 for matter to become relativistic by the neutrino heating, in which 
$\rho$ and  $q^+$ represent the local density and the heating rate, 
  $\tau_{\mathrm{rel}} \equiv \rho c^2/q^+$ is the timescale 
 when the motion of the matter becomes relativistic due to the energy deposition,
  $\tau_{\mathrm{int}} \equiv e_{\mathrm{int}}/q^+$ is the timescale 
 when the neutrino heating is comparable to the internal energy ($e_{\mathrm{int}}$)
 and thus affects the dynamics, 
 $\tau_{\mathrm{hyd}} \equiv X/|v_r|$ indicates the hydrodynamical timescale with 
 $X$ and $v_r$ being the length and radial velocity along the polar axis,
 $\tau_{\mathrm{light}} \equiv X/c$ is the light crossing timescale.
As shown,  $\tau_{\mathrm{int}}$ gets shorter to be several milliseconds near 
around 100 km, where 
 the timescales become most close to $\tau_{\mathrm{hyd}}$. 
This means that rather in the vicinity of the center ($\lesssim 200$ km along the 
 rotational axis), the neutrino heating have 
potential importance to affect the hydrodynamics.
However, it should be noted that it does not directly mean 
that the heated matter could become relativistic. 
In fact, $\tau_{\mathrm{rel}}$ is at least two orders-of-magnitudes larger than 
$\tau_{\mathrm{light}}$.

For making the outflows relativistic, one possible way is to decrease 
$\tau_{\mathrm{rel}}$ by lowering the density of the funnel regions.
Such low density regions would be formed if we continue to follow the 
 dynamics of collapsar in more longer term. Unfortunately however, 
the numerical difficulty of treating such force-free fields prevents us from doing so.
The numerical code specially developed to solve the force-free fields is 
 required (e.g., \citet{mckinney}), which is major undertaking. 
Another way is to increase the heating rate $q^+$.
If the neutrino heating not just from the equatorial plane 
here but also from the whole accretion disk could be included such as by 
 the ray-tracing calculation \citep{birkl}, 
the deposition rates would become larger due to the geometrical effects. 
Dissipative processes such as magnetic-reconnection/Joule heating 
 (e.g., \citet{prog03a}) inside the disk 
would rise the temperature, which could be also good for achieving the 
 higher luminosity. 
Furthermore, general relativistic effects would increase the deposition rate 
in the vicinity of the BH (\citet{salm99,asano1,asano2}), which is also remained to 
 be studied. 

\subsection{General relativistic effects}\label{subsec:kerr}

As mentioned earlier, we have employed the Paczynsky-Wiita potential to mimic 
the Schwarzschild metric. 
With the special relativistic modification, 
this artificial potential is known to be able to 
approximate the general relativistic (GR) motion 
well for the regions $r > 3r_g$ (with $r_g$ being the Schwarzshild radius, 
 e.g., \citet{abram96}, \citet{fukue04}), 
to which we focused on in this study.
  However needless to say, what is needed for collapsars is 
 GR simulations around the Kerr BH 
(e.g., \citet{shiba,nada} and see discussions in \citet{woos06}). 
 
The maximum mass of neutron stars, depending on nuclear equations of state, 
 is estimated to be less than typically 3.0$\Ms$ \citep{latt01,zhang08}. 
  Following this criteria,
the central objects of models with smaller initial angular momenta of J1.0 and J1.5 
 may collapse into the BHs near within 3 s (Figure \ref{time_Mass} (left)). 
In such cases, GR effects very close to the 
inner edge of the accretion disk should be important and their impacts on the MHD 
outflows and the neutrino heating should be investigated. 
It is naturally 
 expected that strong gravity due to GR effects will lead to not only the
 efficient gravitational-wave emission, but also the enhanced 
neutrino emission due to the compression. As recently studied extensively 
 in the context of core-collapse supernovae (e.g., \citet{kota09,kawa09,ott09} 
and references therein), gravitational-wave and neutrino signatures also from 
 collapsars should give us a new observational window to probe the central engine. 
This paper is a prelude before our forthcoming work to clarify
 those aspects, which will be presented elsewhere soon.

\subsection{Numerical Tests}\label{tests}

\begin{figure}[tbd]
\begin{center}
\epsscale{1.0}\plotone{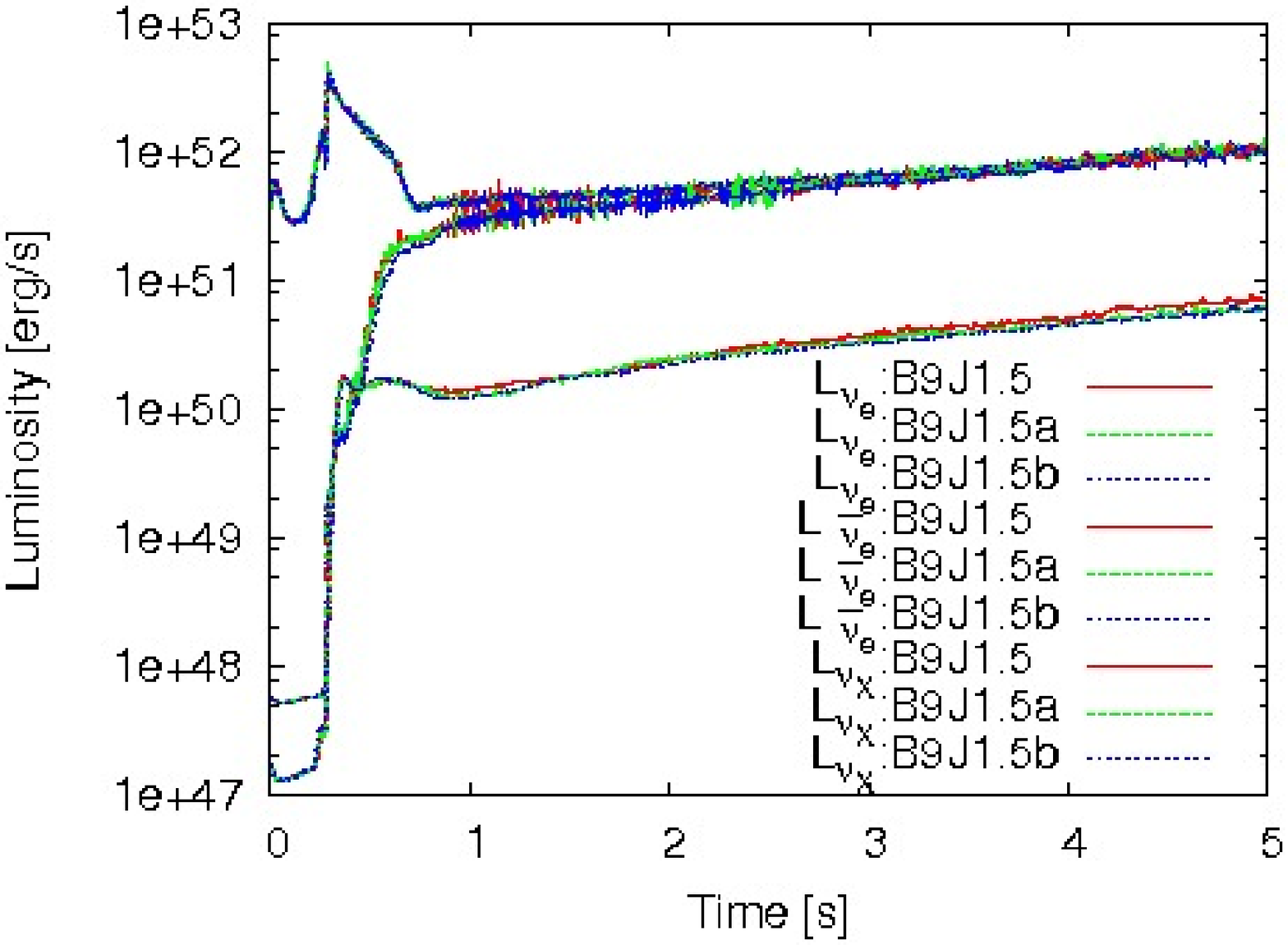}
\caption{Time evolution of neutrino luminosities of 
$\nu_e$ (solid line), ${\bar \nu}_e$ (dashed line), and $\nu_X$ (dotted line) 
 for model B9J1.5.
 Models labeled by ``a'' and ``b'' have doubled mesh 
points in the radial and azimuthal directions than the canonical one of 
300($r$) $\times$ 40($\theta$) mesh points.
Each luminosity coincides well with each other, showing the numerical convergence
 of the obtained results.}
\label{test_L}
\end{center}
\end{figure}

\begin{figure}[tbd]
\begin{center}
\epsscale{1.0}\plotone{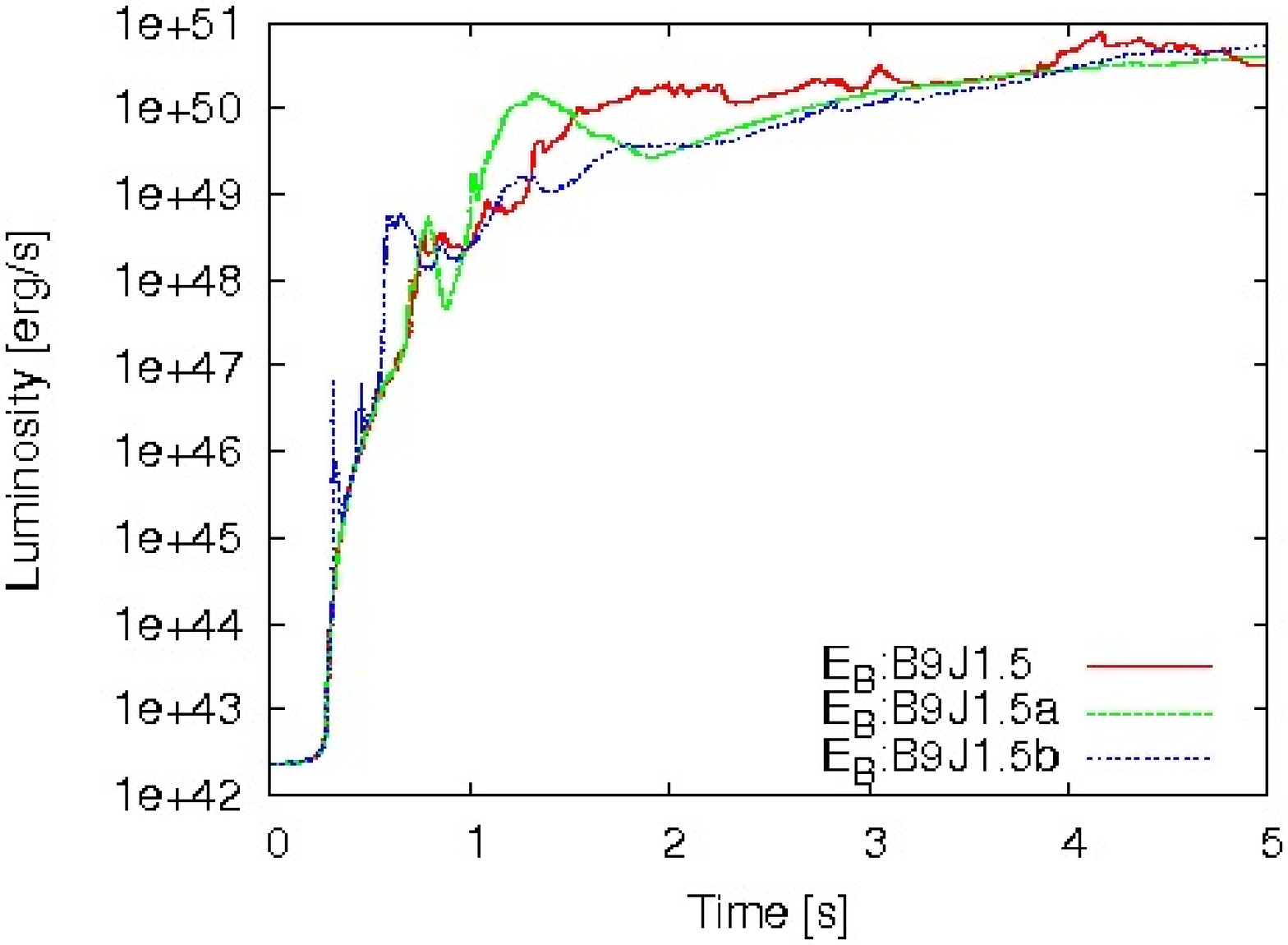}
\caption{Same as Figure \ref{test_L} but for the time evolution of magnetic energy.}
\label{test_Emag}
\end{center}
\end{figure}

Figures \ref{test_L} and \ref{test_Emag} show a convergence of our 
numerical results for different grid resolutions. 
 Figure \ref{test_L} shows an agreement of the neutrino luminosity. 
This means that the evolution of the accretion disk, whose temperature 
and density profiles determine the neutrino luminosity, is numerically
 converged for the different resolution we tested. 
On the other hand, Figure \ref{test_Emag} shows that the magnetic energies 
before $\sim$ 3 sec are rather sensitive to the numerical resolution, although 
the overall trends are similar. We suspect that 
 the discrepancy in the earlier phase should come mainly from the effects of MRI, which 
 are difficult to be captured by our numerical resolution as already mentioned in 
\ref{subsec:magamp}. 

\begin{figure}[tbd]
\begin{center}
\epsscale{1.0}\plotone{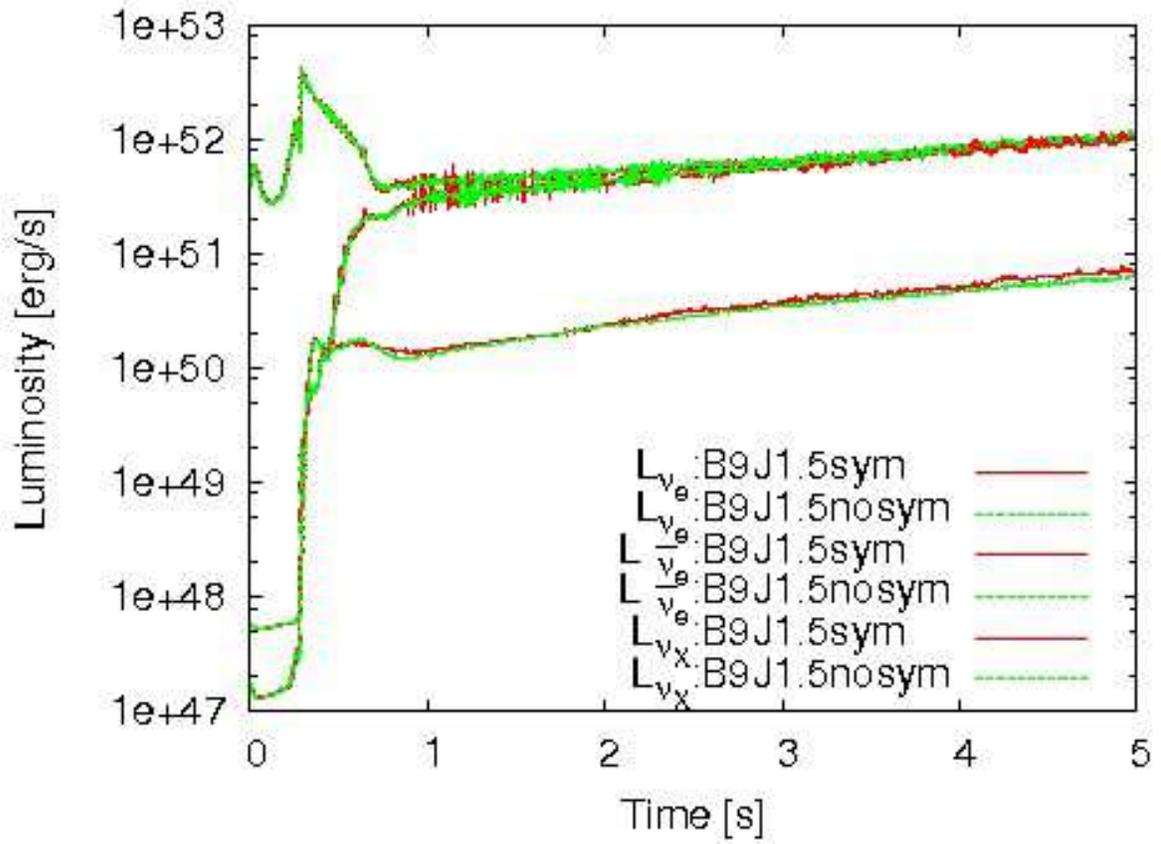}
\caption{Same as Figure \ref{test_L} but with and without equatorial symmetry (indicated by ``sym'' and ``no sym'', respectively). }
\label{time_L_nosym}
\end{center}
\end{figure}

\begin{figure}[tbd]
\begin{center}
\epsscale{1.0}\plotone{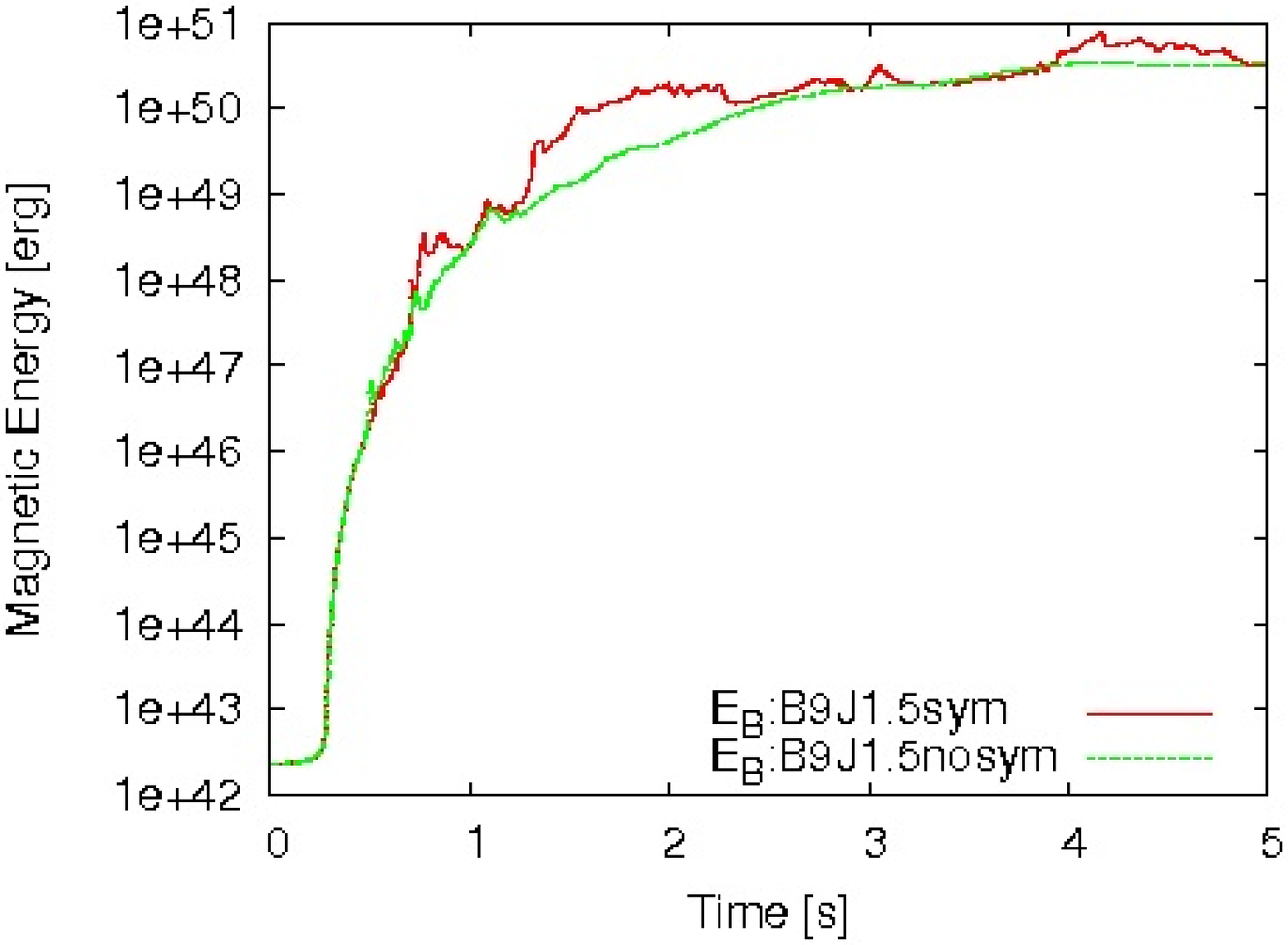}
\caption{Same as Figure \ref{time_L_nosym} but for
 the time evolution of magnetic energy. }\label{time_Emag_nosym}
\end{center}
\end{figure}

Then we discuss the validity of the equatorial symmetry 
assumed in this study. Figure \ref{time_L_nosym} shows that the obtained luminosities 
agree well with each other, however the magnetic energy differs up to two times (Figure 
\ref{time_Emag_nosym}). This could be also due to MRI. 
As mentioned, our numerical resolutions can treat MRI marginally in the sense that 
 it can follow MRI in the late phase when the field strength becomes stronger, 
while it cannot resolve MRI sufficiently 
in the early phase when the field strength is weaker. 
In this sense, the obtained results should give a lower bound for the criterion 
 of the field amplification and the jet formation.
To capture MRI fully is unquestionably important for 
collapsar simulations, but possibly needs the prescription such as by 
the adaptive mesh refinement scheme (e.g., \citet{zhang09})
, which we pose as a future task.



\section{Summary}
In light of the collapsar model of gamma-ray bursts (GRBs), 
 we presented our numerical results of 
 two-dimensional magnetohydrodynamic (MHD) simulations 
of the collapse of rotating massive stars.
 Pushed by recent evolution calculations of
  GRB progenitors, we focused on lower angular momentum of the central 
core than previously assumed.
As for the initial magnetic field strength, we chose to explore relatively smaller 
field strength ($\lesssim 10^{10}$ G), which has been less investigated so far.
By performing special relativistic simulations including 
both realistic equation of state and neutrino cooling, we followed 
a long-term evolution of the slowly rotating collapsars up to $\sim$ 10 s. 
Then we studied how the formation of MHD jets and the properties of accretion disks 
could change with the initial angular momentum and 
 the initial magnetic field strength.
The obtained results can be summarized as follows.

\begin{enumerate}
\item 
Our numerical results show that for the GRB progenitors to function as collapsars, 
there is a critical initial angular momentum: 
${j_{\rm crit} = 1.5 j_{\mathrm{lso}}}$ with $j_{\mathrm{lso}}$
 being the angular momentum of the last stable orbit, below which 
matter is quickly swallowed to the central objects, 
no accretion disks and no MHD outflows are formed. 

\item When larger than the criteria, we find that smaller 
 initial angular momentum leads to more compact accretion disk due to 
compression. It seemed widely to be agreed in previous collapsar simulations 
 that the differential rotation is a primary 
agent to amplify the toroidal fields and the resulting MHD outflows.
 On the other hand, our results show that for relatively slow rotation 
 models, the amplification of the poloidal fields by compression 
 is preconditioned for the amplification of the toroidal fields and the 
 MHD outflows.

\item 
Among the computed models, we find the launch of the MHD jets in the following two ways.
 For models with stronger initial magnetic fields
 ($B_0 \gtrsim 10^{10}$ G),  the gradient of the magnetic pressure perpendicular to the equatorial plane inside the accretion disk 
 can drive the MHD outflow. This outflow makes the strong magnetic explosions 
 like a 'magnetic tower', which we called as type II. For models with weaker 
 initial magnetic fields, the magnetic tower stalls first and the subsequent MHD 
outflow is produced by the accreting material from the equator to the polar region.
 Type I jet is found to be produced when such flow-in materials obtain sufficient 
 magnetic amplifications, due to the strong differential rotation near 
the rotational axis, to overwhelm the ram pressure of the 
accreting material. 

\item
Regardless of type I or II,  the jets can 
 attain only mildly relativistic speeds with 
the explosion energy less than $10^{49} \erg$.
Such events could possibly be related to the X-ray flashes.
 After the passage of the MHD jets, the baryon-poor environments will be  
 left behind. Such polar funnels could provide a favorable cite 
for the subsequent jets, 
which could attain high Lorentz factors pushed by the magnetic outflows and/or 
heated by neutrinos.

\item 
To obtain stronger neutrino energy deposition in the polar funnel 
 heated from the accretion disk, we find that 
 the smaller initial angular momentum is more appropriate.
 This is because the gravitational compression makes 
 the temperature of the disk higher.
 When the accretion disk settles to the quasi-stationary state in their late time 
 evolution, the maximum neutrino luminosity is found to reach $\sim 
10^{52}$ erg/s. Due to the high neutrino opacity inside the disk, the luminosities of 
 $\nu_e$ and $\bar{\nu}_e$ is found to become almost comparable, 
which is advantageous for making the deposition rate larger. 
 Based on an order-of-magnitudes estimation of the energy deposition, 
it is suggested that the neutrino heating could be as efficient 
 as the magnetic mechanism to energetize the outflow. Among the 
computed models here, the model with the initial angular momentum of
 $j_{\rm crit} \sim 1.5j_{\mathrm{lso}}$ and with initial magnetic field strength 
 of $B_0 \gtrsim 10^{10}$ G, provides the most plausible condition for collapsar,
 because such models are appropriate not only for producing the MHD outflows 
 quickly by the magnetic towers,
  but also for obtaining the 
 stronger neutrino heating in the evacuated polar funnel.
 \end{enumerate}

\acknowledgements{S.H. is grateful to T. Kajino for fruitful discussions. 
T.T. and K.K. express thanks to K. Sato and S. Yamada for continuing encouragements.
Numerical computations were in part carried on XT4 and 
general common use computer system at the center for Computational Astrophysics, CfCA, the National Astronomical Observatory of Japan.  This
study was supported in part by the Grants-in-Aid for the Scientific Research 
from the Ministry of Education, Science and Culture of Japan
(Nos. S19104006, 19540309 and 20740150).}

\appendix

\def\thesection{Appendix \Alph{section}}
\def\theequation{\Alph{section}\arabic{equation}}
\section{Definition of energy}
\label{ap:energy}
According to \citet{taki09}, we define special relativistic description of 
local energies as follows, 
\begin{eqnarray}
	e_\mathrm{kin} &=& \rho W\left( W-1\right), \\
	e_\mathrm{int} &=& e W^2 + p \left( W^2-1\right), \\
	e_\mathrm{B^i} &=& \frac{{B_i}^2}{4\pi}\left(1-\frac{1}{2W^2}\right), \\
	e_\mathrm{mag} &=& \sum_{i} e_\mathrm{B^i}, \\
        e_\mathrm{local} &=& e_\mathrm{kin} + e_\mathrm{int} + e_\mathrm{mag} + \Phi_\mathrm{tot}
	\label{def_e}
\end{eqnarray}
where $e_\mathrm{kin}$ is the total kinetic energy, 
$e_\mathrm{rot}$ is the rotational energy, 
$e_\mathrm{int}$ is the internal energy, 
$e_\mathrm{B_i}$ is the magnetic energy of $i$ th component, 
$e_\mathrm{mag}$ is the total magnetic energy, 
and $e_\mathrm{local}$ is the total local energy. 
We use these values to compute the explosion energy of jets 
in \S\ref{sec:results}. 

\section{Special relativistic treatments  for neutrino cooling}\label{leak}

Following the procedure in \citet{mm}, we derive the neutrino cooling rates including special relativistic corrections.
 Neutrino emissivity is often calculated in the rest frame of relativistic fluids
 because radiation is usually isotropic there. 
When converting quantities in the rest frame 
 to the laboratory (lab) frame in which the fluid variables are 
 defined and evolved by the hydrodynamic equations, 
we need to take into account corrections from the Lorentz transformations between
 the two frames.
 
 The cooling rate in the lab frame is calculated as follows,
\begin{equation}
    q^{-} = \int \eta(\Omega,\epsilon) d\epsilon d\Omega,
\end{equation}
by summing up the neutrino emissivity of 
$\eta(\Omega,\epsilon) (\erg\pcmc\psec\pstr)$ of a neutrino energy of $\epsilon$ both over a solid angle $d\Omega$ and over 
 $d\epsilon$ in the lab frame.
 To satisfy the number conservation under the Lorentz transformation, 
 $(\eta/\epsilon) dt dV d\epsilon d\Omega$ is a Lorentz invariant. 
It is noted that the Lorentz transformation can be written in the following simple form
 \citep{mm},
\begin{eqnarray}
    dtdV &=& dt_0dV_0 \\
    d\epsilon &=& \frac{\epsilon}{\epsilon_0}d\epsilon_0 \\
    d\Omega &=& \frac{\epsilon_0^2}{\epsilon^2} d\Omega_0 \\
    \epsilon &=& \gamma \left(1+\beta\mu_0\right) \epsilon_0 \nonumber \\
    &=& \epsilon_0 / [\gamma \left(1-\beta\mu\right)],  \label{lor}
\end{eqnarray}
where $\beta = |\mbox{\boldmath$v$}|/c, \gamma = (1-\beta^2)^{-1/2}$
, and $\mu = \mbox{\boldmath$v$}\cdot\mbox{\boldmath$n$}/|\mbox{\boldmath$v$}|$ 
where $\mbox{\boldmath$n$}$ is the unit vector of 
the direction of the emitted neutrino. 
Here the subscript 0 denotes the variables which are measured in the rest frame. 
Then, we obtain the transformation of the emissivity, 
\begin{equation}
    \eta(\mu,\epsilon) = \frac{\epsilon^2}{\epsilon^2_0}\eta_0(\mu_0,\epsilon_0). 
\end{equation} 
When the emission is isotropic in the rest frame, namely as
\begin{equation}
    \eta_0(\mu_0,\epsilon_0) = \zeta_0(\epsilon_0), 
\end{equation} 	
where $\zeta_0(\epsilon_0) [\erg\pcmc\psec]$ is the direction-averaged 
emissivity, the cooling rate in the lab frame can be given as, 
\begin{eqnarray}
    q^{-} &=& \int \eta(\Omega,\epsilon)d\epsilon d\Omega \nonumber \\ 
    &=&\int \frac{\epsilon}{\epsilon_0}\eta_0(\mu_0,\epsilon_0) 
    d\epsilon_0 d\Omega_0 \nonumber \\
    &=& \int \left[ \int \gamma \left(1+\beta\mu_0\right) d\Omega_0 \right] 
    \zeta_0(\epsilon_0)d\epsilon_0 \nonumber \\ 
    &=& 4 \pi \gamma \int \zeta_0(\epsilon_0)d\epsilon_0. \nonumber \\ 
    &=& \gamma q^{-}_0.\label{eq:cool}
\end{eqnarray}
This equation means that the cooling rate becomes $\gamma$ times larger  
 than the one in the rest frame with the special relativistic effect. 
It is noted that this relation holds when 
$\eta_0(\mu_0,\epsilon_0)$ is isotropic in the rest frame, such as 
inside the accretion disk investigated here, which is very opaque to neutrinos and thus the 
 radiation there is well approximated to be isotropic (see discussions in 
 section 4.3). 

Now we move on to consider the corrections to the opacity.
Again from the number conservation with the Lorentz transformation of 
the absorption coefficient $\chi(\Omega,\nu)\ (\erg\pcmc\psec\pstr)$, 
$(\chi/\epsilon) dt dV d\epsilon d\Omega$ is a Lorentz invariant. 
Therefore, with the same approach in the case of the cooling, 
the absorption coefficient in the lab frame becomes,
\begin{eqnarray}
    \chi(\mu,\epsilon) &=& \frac{\epsilon_0}{\epsilon} {\chi}_0(\mu_0,\epsilon_0) \nonumber \\
    &=& \gamma \left(1-\beta\mu\right) \chi_0(\mu_0,\epsilon_0).
\end{eqnarray}
Then the opacity can be calculated as, 
\begin{eqnarray}
    d\tau &=& \chi(\mu,\epsilon) ds \nonumber \\
    &=& \gamma \left(1-\beta\mu\right) \chi_0(\mu_0,\epsilon_0) ds \nonumber \\
    &=& \gamma \left(1-\beta\mu\right) d\tau_0, \label{eq:opacity}
\end{eqnarray}
where $ds$ is the spatial distance in the lab frame. Here  
$(\gamma \left(1-\beta\mu\right))$ reflects the special relativistic effect. 

With the local cooling rate (Eq. (\ref{eq:cool})) and the opacity 
(Eq. (\ref{eq:opacity})), 
 we evaluated the effective cooling rate appeared in Eq. (\ref{eq:ene_consv}) as 
\begin{eqnarray}
    {\cal L}_\nu &=& q^- \exp \left[-\int d\tau \right] \nonumber \\
    &=& \gamma q^-_0 \exp \left[-\int \gamma \left(1-\beta\mu\right) d\tau_0 \right], 
\end{eqnarray}
where the integration is performed along the radial direction for simplicity. 

Finally, the reaction rate is also modified in the lab frame 
 by the time contraction. 
The time evolution of $Y_e$ is then given by 
\begin{eqnarray}
    dY_e &=& \Gamma_0 dt_0 \nonumber \\
    &=& \Gamma_0 \gamma dt, 
\end{eqnarray}
where $\Gamma_0$ represents the local neutrino reaction rates.
Apparently, this modification could be important in the regions where 
 the motion becomes (special)relativistic such as the inner most region of the disk 
or inside of the relativistic outflow. 

\section{Special relativistic modification for neutrino pair annihilation}
\label{ap:srheat}
Employing the same procedure as in the previous section, we here derive the 
 heating rate via neutrino pair annihilation ($\nu + {\bar \nu} \rightarrow e^{-}+ e^{+}$) with special relativistic corrections.

By putting the Lorentz transformations described in Eq.(\ref{lor}) 
into the heating rate derived by previous studies \citep{salm99,asano1}, 
the heating rate in the lab frame is given by 
\begin{eqnarray}
    \frac{dq_{\nu {\bar{\nu}}}^+(\mbox{\boldmath$r$})}{dt dV}&=&
	2 c K G_{\rm F}^2 
    \int
    d\theta_{\nu} d \phi_{\nu}  d\theta_{\bar{\nu}} d \phi_{\bar{\nu}}
    d\epsilon_{\nu} d \epsilon_{\bar{\nu}}
    \epsilon_{\nu}^2 \epsilon_{\bar{\nu}}^2 
    (\epsilon_\nu + \epsilon_{\bar{\nu}} ) f_{\nu}(\mbox{\boldmath$r$}, \mbox{\boldmath$p$}_\nu)
    f_{\bar{\nu}}(\mbox{\boldmath$r$}, \mbox{\boldmath$p$}_{\bar{\nu}})
    \nonumber \\
    && \times \left[ 1-\sin{\theta_{\nu}} \sin{\theta_{\bar{\nu}}}
        \cos{(\varphi_{\nu}-\varphi_{\bar{\nu}})}
        -\cos{\theta_{\nu}} \cos{\theta_{\bar{\nu}}}
    \right]^2
    \sin \theta_{\nu}\sin \theta_{\bar{\nu}} \nonumber \\
	&=& 2 c K G_{\rm F}^2 
    \int
    d\theta_{\nu} d \phi_{\nu}  d\theta_{\bar{\nu}} d \phi_{\bar{\nu}} \nonumber \\
    && \times [ SR_\nu^5 (\Omega_\nu) SR_{\bar{\nu}}^4 (\Omega_{\bar{\nu}}) S_{\nu,0}(\mbox{\boldmath$r$},\Omega_\nu) N_{\bar{\nu},0}(\mbox{\boldmath$r$},\Omega_{\bar{\nu}}) \nonumber \\
         && \quad + SR_\nu^4 (\Omega_\nu) SR_{\bar{\nu}}^5 (\Omega_{\bar{\nu}}) N_{\nu,0}(\mbox{\boldmath$r$},\Omega_\nu) S_{\bar{\nu},0}(\mbox{\boldmath$r$},\Omega_{\bar{\nu}})  ]
    \nonumber \\
    && \times \left[ 1-\sin{\theta_{\nu}} \sin{\theta_{\bar{\nu}}}
        \cos{(\varphi_{\nu}-\varphi_{\bar{\nu}})}
        -\cos{\theta_{\nu}} \cos{\theta_{\bar{\nu}}}
    \right]^2
    \sin \theta_{\nu}\sin \theta_{\bar{\nu}},
 \label{eq:heat_ann}
\end{eqnarray}
where 
\begin{eqnarray}
	SR_\nu(\mbox{\boldmath$r$},\Omega_\nu) &=& \epsilon_\nu / \epsilon_{\nu,0}\nonumber \\
        &=& 1/[\gamma_\nu(1-\mu_{\nu}\beta_{\nu})], \label{eq:heatdef_start}\\
	\beta_\nu   &=& \frac{|\mbox{\boldmath$v$}_\nu|}{c}, \\ 
	\gamma_\nu  &=& \frac{1}{\sqrt{1-\beta_{\nu}^2}}, \\
	\mu_{\nu}   &=& \frac{\mbox{\boldmath$n$}_\nu\cdot\mbox{\boldmath$v$}_\nu}{|\mbox{\boldmath$v$}_\nu|} \\
	\mbox{\boldmath$n$}_\nu        &=& \frac{\mbox{\boldmath$p$}_\nu}{|\mbox{\boldmath$p$}_\nu|} 
        = (\sin \theta_\nu \cos \phi_\nu,\  
        \sin \theta_\nu \sin \phi_\nu, \ 
        \cos \theta_\nu), \\
	S_{\nu,0}(\mbox{\boldmath$r$},\Omega_\nu)  &=& \int \epsilon_{\nu, 0}^4 f_{\nu,0}(\mbox{\boldmath$r$}_{\nu, 0}, \mbox{\boldmath$p$}_{\nu,0}) d\epsilon_{\nu,0}, \\	
	N_{\nu,0}(\mbox{\boldmath$r$},\Omega_\nu)  &=& \int \epsilon_{\nu, 0}^3 f_{\nu,0}(\mbox{\boldmath$r$}_{\nu, 0}, \mbox{\boldmath$p$}_{\nu,0}) d\epsilon_{\nu, 0}, \label{eq:heatdef_end}
\end{eqnarray}
and $\mbox{\boldmath$r$}_\nu, \mbox{\boldmath$v$}_\nu$, 
$\mbox{\boldmath$p$}_\nu$, $f_\nu$ denotes 
the position of the neutrino source, 
the velocity of fluid at the neutrino source, the momentum vector of neutrino, 
and the distribution function of neutrino. 
Definitions are same for the anti-electron type neutrino by changing the 
notation $\nu$ to ${\bar \nu}$ from Eq.(\ref{eq:heatdef_start}) to 
Eq.(\ref{eq:heatdef_end}). 
 Subscript 0 again denotes variables which are measured in the rest frame of 
 the neutrino source. Here the neutrino source indicates the accretion disk.
The neutrino number flux along its path to the target is assumed to 
 be conserved for simplicity as $f_{\nu}(\mbox{\boldmath$r$}, \mbox{\boldmath$p$}_\nu)
=  f_\nu(\mbox{\boldmath$r$}_{\nu, 0}, \mbox{\boldmath$p$}_{\nu, 0})$. 
In Eq.(\ref{eq:heat_ann}), 
the factor $SR_\nu$ reflects the special relativistic modification to 
the heating rate. 


The neutrino distribution function inside the accretion is well approximated by the 
 one in the $\beta$ equilibrium as, 
\begin{eqnarray}
    f_\nu(\mbox{\boldmath$r$}_{\nu, 0}, \mbox{\boldmath$p$}_{\nu, 0})
    &=& \frac{1}{(hc)^3} \frac{dn_0}{d\epsilon_0 d\Omega_0 dt_0 dV_0} \nonumber \\
	&=&\frac{1}{(hc)^3}\frac{1}{\exp(\epsilon_{\nu,0} / kT_{\nu,0}-\eta_{\nu,0}) + 1}, 
\end{eqnarray}
where $T_{\nu,0}$, $\eta_{\nu,0}$ is the temperature and 
 degeneracy parameter of neutrino in the rest frame, respectively. 
We approximate $T_{\nu,0}$ to be equal to 
the temperature of fluid $T(\mbox{\boldmath$r$}_\nu)$. 
Then, $S$ and $N$ in Eq. (C1) can be expressed by the Fermi integrals 
${\cal F}_k$ as 
\begin{eqnarray}
    {\cal F}_k(y) &\equiv& 
    \int \limits_0^\infty
    \frac{x^k}{\exp(x-y) + 1}
    dx, \\
    S_{\nu,0}(\mbox{\boldmath$r$},\Omega) &=& \frac{(kT(\mbox{\boldmath$r$}_\nu))^5}{(hc)^3}{\cal F}_4(\eta_{\nu,0}), \\
    N_{\nu,0}(\mbox{\boldmath$r$},\Omega) &=& \frac{(kT(\mbox{\boldmath$r$}_\nu))^4}{(hc)^3}{\cal F}_3(\eta_{\nu,0}). 
\end{eqnarray}
With these modifications, 
we calculated the heating rate in section \ref{subsec:heat}.

\bibliographystyle{apj}
\bibliography{ms}

\end{document}